\documentclass[reprint,english, amsmath,amssymb, aps, prl,floatfix]{revtex4-2}
\usepackage{braket}
\usepackage{cleveref}
\usepackage{float}
\usepackage{tabularx}
\usepackage{soul}
\usepackage{array}
\usepackage{chngcntr}
\usepackage[normalem]{ulem}
\usepackage{graphicx}
\usepackage{dcolumn}
\usepackage{bm}
\usepackage[table]{xcolor}

\counterwithout{equation}{section}

\usepackage[english]{babel}
\definecolor{light_gray}{rgb}{0.95, 0.95, 0.95}

\begin{document}

\preprint{APS/123-QED}

\title{Fractional quantum Hall edge polaritons}

\author{Lucas Winter}
\email{lucas.s.winter@icloud.com}
\author{Oded Zilberberg}%
\affiliation{%
 Fachbereich Physik, Universit\"at Konstanz, DE-78457 Konstanz, Germany
}%

\date{\today}

\begin{abstract}
It is commonly believed that light cannot couple to the collective excitations of the fractional quantum Hall effect (FQHE). This assumption relies on Kohn's theorem that states that electron-electron interactions decouple from homogeneous electromagnetic fields due to Galilean invariance. 
Here, we demonstrate that light-matter coupling beyond the dipole approximation circumvents Kohn's theorem, and enables the coupling of cavity photons to the plasmonic edge modes of the FQHE. 
We derive the coupling using the FQHE bulk-boundary correspondence and predict the formation of experimentally detectable plasmon polaritons. In conjunction with recent experiments, we find that a single cavity mode leaves the system's topological protection intact. Interestingly, however, a multimode cavity mediates plasmon backscattering and effectively transforms the edges of the 2D FQHE into a 1D wire. Such cavity-meditated nonlocal backscattering bodes the breakdown of the topological protection in the regime of ultra-strong photon-plasmon coupling. Our analytical framework and findings pave the way for investigating the topological order of the FQHE via optical spectroscopic probes as well as provide new opportunities to control FQHE edge excitations using light. 
\end{abstract}

\maketitle


\begin{figure}
    \centering
    \includegraphics[width = 0.99\columnwidth]{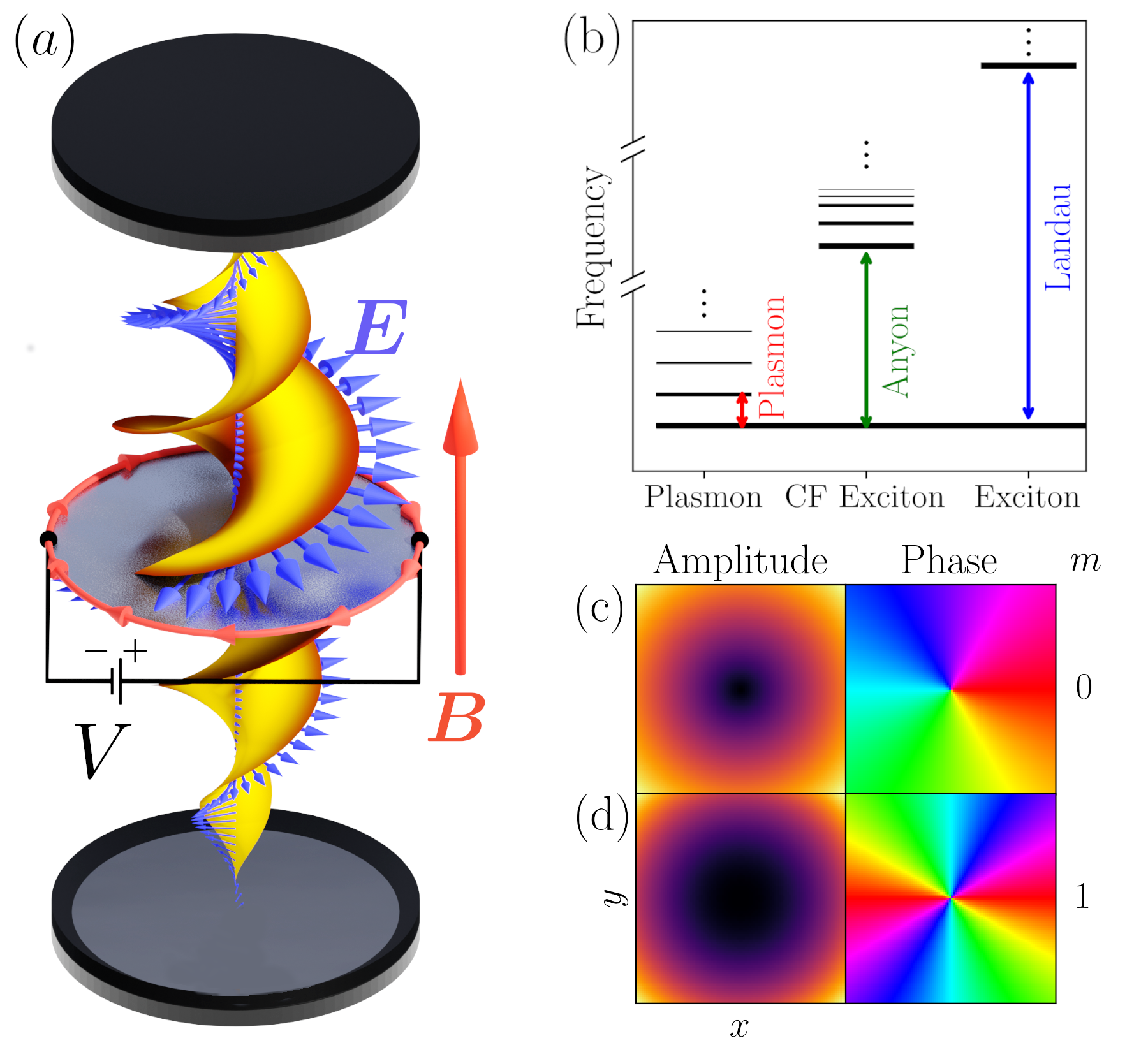}
    \caption{\textit{System}. (a) A 2D Hall disc is positioned inside an optical cavity, cf. Eq.~\eqref{eq:HamiltonianCoulomb}. The Hall disc is subjected to a perpendicular magnetic field $\boldsymbol{B}$ and is coupled to a cavity electric field $\boldsymbol{E}$, which can be decomposed into angular momentum modes $l$, cf.~Eq.~\eqref{eq:EM-vector}. We show a single circularly polarized basis mode $\boldsymbol E_l$ (blue arrows) with a helical wavefront (orange surface). A voltage bias $V$ can be applied across the edge of the Hall bar.  Here we consider mesoscopic systems with \(\nu=1/3\) and \(n=10^{11}\,\mathrm{cm}^{-2}\): \(B\approx12.4\,\mathrm{T}\), \(\ell_B\approx7\,\mathrm{nm}\);
circumference \(L=100\,\mu\mathrm{m} \); }
(b) The hierarchy of energy scales of the Hall effect, corresponding to potential excitation by the cavity, ranges from small to large energies: FQHE edge plasmons (red), FQHE bulk excitations (green), and IQHE Landau-level spacing (blue). 
(c) and (d) Examples of the cavity basis mode’s amplitude and phase in the plane of the Hall disc for different OAM (Orbital Angular Momentum) quanta $m$, cf. Eq.~\eqref{eq:EM-vector}.
    \label{fig:setup}
\end{figure}

One of the most fascinating effects in condensed matter physics is the emergence of collective excitations arising from strong interactions between particles~\cite{lerner2002strongly,dagotto2005complexity}. A principal setting for the formation of such exotic quasiparticles is the fractional quantum Hall effect (FQHE)~\cite{tsui_two-dimensional_1982,frohlich1993gauge, stormer1999fractional,k_jain_jainendra_composite_2007,stern2008anyons,frohlich2023gauge}. Here, electrons with charge $e$ and mass $m$ are confined to move in 2D in the presence of a perpendicular magnetic field $B$. The competition between their kinetic energy and the magnetic field suppresses the dispersion, and leads to cyclotron motion at frequency $\omega_c = eB/m$. As a result, a  highly-degenerate Landau-level spectrum with gaps $\hbar \omega_c$ manifests. In this setting, electron-electron interactions can dominate and open a many-body gap at specific fractional fillings of the Landau levels, where new quasiparticles emerge. These collective excitations are dubbed anyons and are predicted to obey fractional exchange statistics, i.e., they are neither bosonics nor fermionic~\cite{stern2008anyons,campagnano2012hanbury,campagnano_hanbury_2013}, and can even be used for topological quantum computation~\cite{nayak2008non,zilberberg2008controlled}. At so-called Laughlin fillings $\nu = 1/(2p + 1)$ with $p \in \mathbb N$~\cite{tsui_two-dimensional_1982, laughlin_anomalous_1983, stormer_nobel_1999}, the resulting FQHE is well-described using composite fermions (CFs) that support anyonic bulk excitations with energy scales on the order of the Coulomb energy $\sim e^2/(\epsilon l_B)$, where $\epsilon$ is the dielectric constant and $l_B$ is the magnetic length~\cite{k_jain_jainendra_composite_2007}.

Strong light-matter coupling in optical cavities can also lead to interesting collective excitations and effects~\cite{fleischhauer2005electromagnetically,carusotto2013quantum,bloch2022non}. For example, it allows for novel topological phases populated by light-matter hybrid particles known as polaritons~\cite{ozawa2019topological,solnyshkov2021microcavity}. In relation to the quantum Hall effect, coupling to light was used both for probing and for modifying the topological order using cavities~\cite{schlawin_cavity_2022, frisk_kockum_ultrastrong_2019, forn-diaz_ultrastrong_2019, jaako_ultrastrong-coupling_2016, ruggenthaler_quantum-electrodynamical_2018, garcia-vidal_manipulating_2021, sentef_cavity_2018, kirton_introduction_2019}. In the non-interacting limit, and using very high frequencies, transitions between Landau levels were studied, leading to the formation of Landau polaritons~\cite{hagenmuller_ultrastrong_2010, scalari_ultrastrong_2012, keller_landau_2020, li_vacuum_2018}, and modification of the quantized electron transport of the integer quantum Hall  effect (IQHE)~\cite{bartolo_vacuum-dressed_2018, paravicini-bagliani_magneto-transport_2019, ciuti_cavity-mediated_2021}. In essence, cavity vacuum fluctuations tend to expunge the quantized Hall transport in the IQHE~\cite{appugliese_breakdown_2022, ciuti_cavity-mediated_2021, rokaj_quantum_2019, rokaj_polaritonic_2022, rokaj_topological_2023}. Moving to the FQHE,  its topological order can be probed indirectly by Landau polaritons~\cite{knuppel_nonlinear_2019, knuppel_two-dimensional_2021}. Furthermore, the induction of bulk anyonic excitations by locally inserting angular momentum with electric fields has been proposed~\cite{gras_optical_2018}. However, in experiments, the FQHE shows significant resilience to such coupling~\cite{appugliese_breakdown_2022}. This resilience stems from Kohn's theorem, which states that a homogeneous electromagnetic field only modifies the center of motion dynamics while leaving electron-electron correlations intact ~\cite{kohn_cyclotron_1961, k_jain_jainendra_composite_2007, maag_coherent_2016,appugliese_breakdown_2022, rokaj_topological_2023}. For the same reason, the topological order seems to be protected from cavity vacuum fluctuation~\cite{appugliese_breakdown_2022, rokaj_topological_2023}.

In realistic scenarios, Hall systems are finite. Due to the so-called bulk-boundary correspondence, in the non-interacting limit, chiral electronic edge modes appear in the gaps between the Landau levels, which carry the IQHE  current~\cite{laughlin_quantized_1981,halperin_quantized_1982}.  Crucial to this work, a similar bulk-boundary correspondence manifests for the CFs, and carries the fractionally-quantized current of the FQHE~\cite{stern2008anyons}. At the edge, we then expect chiral collective bosonic excitations (plasmonic modes), which fill the gaps between the effective CF Landau levels~\cite{wen_chiral_1990, wen_theory_1992}.

In this work, we derive a model for the coupling between cavity light and the FQHE.
 Our theory breaks Kohn's theorem by considering  spatially varying mode profiles beyond the dipole approximation. Our derivation harnesses the bulk-boundary correspondence and implies strong coupling with the system's plasmonic edge modes. As a result, experimentally-observable edge polaritons appear. 
Correspondingly, the cavity introduces backscattering between the edge modes on the opposite sides of the system, which  can impact the Hall conductance. 
In agreement with experiments ~\cite{appugliese_breakdown_2022}, within our model, a  homogeneous cavity mode is insufficient to disturb the quantized Hall conductance. Crucially, however, we find that an inhomogeneous multi mode cavity introduces numerous scattering channels and removes the topological protection. 
Our predictions offer a variety of observations that are within experimental reach. At the same time, our methodology can be applied to a plethora of other topological systems in mesoscopics, as well as to quantum simulators. 

We consider an optical cavity coupled to a spin polarized 2D electron gas (2DEG) in the FQHE regime [see Fig.~\ref{fig:setup}(a)]
\begin{align}
 \resizebox{0.9\hsize}{!}{$
    H_C =  \frac 1{2m_e}\sum_j (\boldsymbol \pi_j - e\boldsymbol A(\boldsymbol{r}_j))^2  + V_{\mathrm{int}} + V_{\mathrm{conf}} + H_{\mathrm{cav}}\, ,  $}\label{eq:HamiltonianCoulomb}
\end{align}
where $\boldsymbol{\pi}_j = \boldsymbol{p}_j - e B x_j\boldsymbol{\hat e}_y$ is the canonical momentum of electron $j$ moving in the plane with momentum $\boldsymbol{p}_j$ in the presence of a perpendicular magnetic field $B$. The latter is written in Landau's gauge. The light-matter coupling involves the cavity vector potential $\boldsymbol{A}(\boldsymbol{r})$ using standard minimal coupling~\cite{k_jain_jainendra_composite_2007}.  We allow $\boldsymbol A(\boldsymbol{r})$ to vary arbitrarily in space, to accommodate any electromagnetic environment. The terms $V_{\mathrm{int}}$ and $V_{\mathrm{conf}}$ are the electrons' Coulomb interaction and confinement potentials, respectively. The term $H_{\mathrm{cav}}=\sum_{l\neq 0} \hbar\omega_{c,l} a_l^\dagger a_l$ describes a multimode cavity with mode frequencies $\omega_{c,l}$ and quantized annihilation operators $a_l$.
In Fig.~\ref{fig:setup}(b), we summarize the hierarchy of energy scales of the FQHE, cf.~Eq.~\eqref{eq:HamiltonianCoulomb} in the absence of the cavity, and the discussion in the introduction.

For simplicity, we assume the Hall system to be rotationally symmetric around the cavity axis. This implies that the total electron angular momentum is conserved. As the angular momentum $l$ is a good quantum number for the electrons, we write the electromagnetic field also in terms of angular momentum modes $l$.  Radially, we use a basis of Hypergeometric-Gaussian modes ~\cite{kotlyar_hypergeometric_2007, karimi_hypergeometric-gaussian_2007}, and assume, in the main text, a broad beam waist relative to the Hall system extent~\cite{supmat}
\begin{align}
\boldsymbol{A} = \sum_{l\neq 0} \boldsymbol{A}_l =\sum_{l\neq 0} A_l(\boldsymbol{\epsilon} z^{|l|-1} \alpha_l^{\phantom{\dagger}} +  \boldsymbol{\epsilon}^* (z^*)^{|l|-1} \alpha_l^\dagger) \, ,
\label{eq:EM-vector}
\end{align}
where $\boldsymbol{\epsilon} =  (1,i, 0)^T/\sqrt{2}$ is the circular polarization vector
and $z = (x + iy)/R$ encodes the position in the 2DEG using a complex coordinate, where $R$ denotes the radius of the disc. The bosonic operators $\alpha_l = a_l \Theta(l) + a_l^\dagger \Theta(-l)$ and $\alpha_l^\dagger = a_l^\dagger \Theta(l) + a_l \Theta(-l)$ create or annihilate a photon with angular momentum $l$. The total photon angular momentum consists of one \textit{quantum spin angular momentum} and $m=l-1$ quanta of \textit{orbital angular momentum} (OAM)~\cite{padgett_lights_2004, forbes_orbital_2021}, see Figs.~\ref{fig:setup}(c) and (d). 
Modes with $|l|>1$ are inhomogeneous and break Kohn's theorem.
Note that the coupling of any 3D cavity field to a 2DEG can be written in the form of Eq.~\eqref{eq:EM-vector} by applying a suitable gauge transform~\cite{supmat}.

 In the following, we develop a,   relevant for DC transport, low-energy effective theory focused on coupling to the plasmonic edge excitations. We truncate matrix elements that couple to high energies, cf.~Fig.~\ref{fig:setup}(b), i.e., our theory is valid for energies 
below the many-body gap. Our model~\eqref{eq:HamiltonianCoulomb}, however, is written in the Coulomb gauge, which is less robust to level truncation than the dipole gauge 
~\cite{de_bernardis_breakdown_2018, di_stefano_resolution_2019}.
Hence, we switch to the dipole gauge using the unitary transformation $H_D = U H_C U^\dagger$ with $U = \prod_j\exp(-ie\sum_{l\neq 0} \boldsymbol{A_l}(\boldsymbol r_j) \cdot \boldsymbol{r}_j/(|l|\hbar))$~\cite{jackson_lorenz_2002, de_bernardis_breakdown_2018}
\begin{align}
\resizebox{0.9\hsize}{!}{$\displaystyle{ 
   H_D =\sum_j \bigg[\frac{\boldsymbol\pi_j^2 }{2m_e}\boldsymbol + \sum_{l\neq 0} \frac{e}{|l|} \ \boldsymbol E_l(\boldsymbol{r}_j) \cdot \boldsymbol r_j\bigg] +  V_{\mathrm{int}} + V_{\mathrm{conf}}+ H_{\mathrm{cav}} \,. }$} \label{eq:DipoleGauge}
\end{align}
 Here,   $\boldsymbol{E_l}\cdot \boldsymbol{r} = iA_l\omega_{c,l}(p_l \alpha_l - p_l^* \alpha_l^\dagger)$ is the dipole coupling of the $l$ component of the electric field, and $p_l = \sum_j z_j^l$ is a symmetric polynomial in the electron coordinates $\{z_j\}$.  We omit the dipole self-energy, which only renormalizes the parameters of our theory and vanishes in the thermodynamic limit~\cite{supmat}.

\begin{figure}[!t]
    \centering
    \includegraphics[width = 0.98\columnwidth]{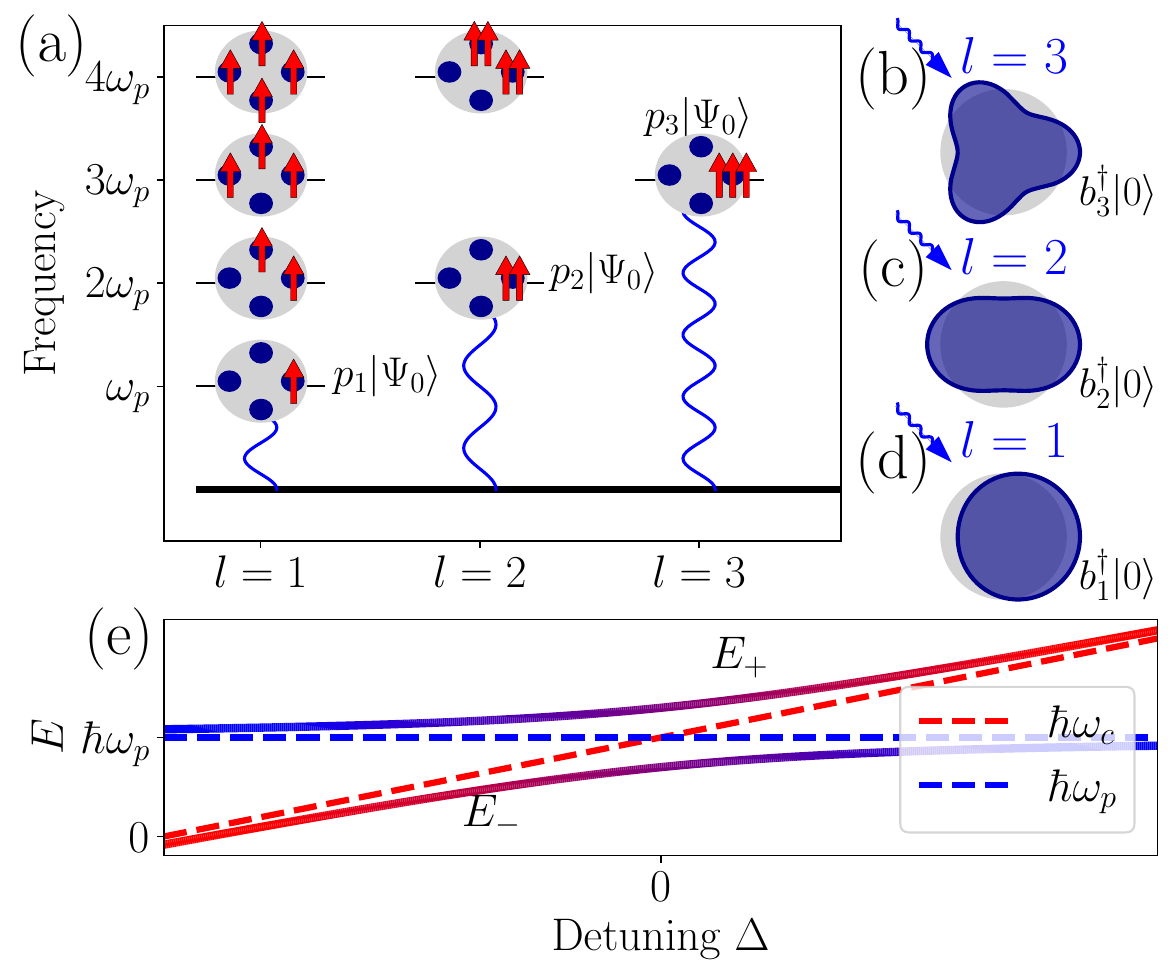}
    \caption{ \textit{Light-matter coupling to the FQHE and formation of edge plasmon-polaritons.} (a) Low-energy spectrum of the FQH disc (with 4 electrons for illustration). Excited states are obtained by attaching one or more angular momentum quanta (red arrows) to electrons (blue circles). A single angular momentum quanta increases the energy by $\hbar \omega_p$ units. Absorbed photons attach or remove angular momentum quanta from single electrons, via optical transitions (squiggly lines) into states $p_l\ket{\Psi_0}$, where $l = 1,2,3$ are marked.
    (b), (c) and (d) According to the bulk-boundary correspondence, bulk states with added angular momentum $p_l \ket{\Psi_0}$ manifest as edge plasmons $  b_l^\dagger \ket 0$.  Photons with angular momentum $l$ (squiggly line) can create plasmons at wavevector $k_l = 2\pi l/L$.
    (e) FQHE edge excitations spectrum in a single-mode cavity field shows plasmon-polariton anticrossing with cavity detuning [cf.~Eq.~\eqref{eq:edgeHamiltonian}]. Dashed lines indicate the spectrum without light-matter interaction. }
    \label{fig:figure2}
\end{figure}

When the FQHE system absorbs a photon $a_l^\dagger$ [see Fig. \ref{fig:figure2}(a)], angular momentum conservation dictates a transition to an excited state, 
with more angular momentum than the Laughlin ground state. We express the Hamiltonian~\eqref{eq:DipoleGauge} in the basis of many-body angular momentum states. Increasing the angular momentum of one electron by $l$ units corresponds to multiplying the ground state with the symmetric polynomial $p_l \ket{\Psi_0}$~\cite{laughlin_anomalous_1983,duncan_hierarchy_1990, simon_wavefunctionology_2020, hansson_conformal_2007, hansson_quantum_2017}, where $p_l = \sum_j z_j^l$. A general excited state can then be written as $\ket{\Psi_{\boldsymbol{\lambda}}} = \prod_l p_n^{\lambda_l} \ket{\Psi_0}$. Each quantum of angular momentum adds energy $\hbar \omega_p$, where $\omega_p$ is the plasmon frequency, see Fig.~\ref{fig:figure2}(a).
The matrix elements of the light-matter coupling in the many-body basis $\ket{\Psi_{\boldsymbol{\lambda}}}$ therefore read 
\begin{equation}
    \label{eq:word}
    \resizebox{0.9\hsize}{!}{%
    $
    \begin{split}
        \braket{\Psi_{ \boldsymbol{\lambda}}|H_{\mathrm{int}}|\Psi_{\boldsymbol{\mu}}}&= \int \prod dz_j \Psi^*_{\boldsymbol \lambda}(\{z_j\})  H_{\mathrm{int}} \Psi_{\boldsymbol \mu}(\{z_j\})  \
\\ &= i\sum_{l\neq 0}\frac{eE_l}{|l|} \bigg(\braket{\Psi_{\boldsymbol{\lambda}} |p_{|l|}| \Psi_{\boldsymbol{\mu}}}\alpha_l^{\phantom{\dagger}} - \braket{\Psi_{\boldsymbol{\lambda}}| p_{|l|}| \Psi_{\boldsymbol{\mu}}}^*\alpha_l^\dagger \bigg) %
    \end{split} 
    $
}
\end{equation}

\noindent with $E_l = A_l \omega_{c,l}$.  Note that the inner products are  evaluated by taking real-space integrals over the many-body Laughlin wavefunction~\cite{supmat}. In our case, we can use the fact that $\braket{\Psi_{\boldsymbol{\lambda}}}$ form an orthogonal basis$\braket{\Psi_{\boldsymbol\lambda}|p_l|\Psi_{\boldsymbol\mu}} \propto+ \delta(\lambda_{l} - \mu_{l} + 1)\prod_{n\neq l} \delta(\lambda_n - \mu_n)$ to simplify the calculation~\cite{simon_wavefunctionology_2020}.  This implies that an absorbed photon $a_l^\dagger$ attaches $l$ quanta of angular momentum to one electron, and induces a transition into an excited state, see Fig.~\ref{fig:figure2}(a).    %

In the thermodynamic limit, the bulk-edge correspondence specifically states that the inner product between bulk wavefunctions is directly related with the inner product between edge states $\braket{\Psi_{\boldsymbol{\lambda}}|\Psi_{\boldsymbol{\mu}}} = \sqrt{\nu l}\  L/(2\pi) \braket{\boldsymbol{\mu}|\boldsymbol{\lambda}}$~\cite{wen_chiral_1990, simon_wavefunctionology_2020}, where $\nu$ is the filling factor $\nu = n_{2D}h/(eB)$ corresponding to electron density $n_{2D}$, and  $L$ is the edge circumference.  Note that in small systems corrections to the inner product relation will appear. Crucial to this work, the bulk-edge correspondence implies that the excited state $p_l\ket{\Psi_0}$  maps to a plasmonic surface wave at the edge $b_l^\dagger \ket{0}$, where $b_l^\dagger$ is a bosonic creation operator acting on an edge state; see Figs.~\ref{fig:figure2}(b)-(d). Each plasmonic excitation is defined by its momentum $\hbar k_l = \hbar \frac{2\pi}{L}l$ and distinct energy $\hbar \omega_p l = \hbar v k_l$, with edge drift velocity $v$. The collective of all plasmonic excitations forms a chiral Luttinger liquid~\cite{wen_chiral_1990}. By using the bulk-edge relations, we evaluate the matrix elements and obtain an effective cavity-plasmon Hamiltonian
\begin{align}
   \resizebox{0.9\hsize}{!}{$\displaystyle{ H_L = \sum_{l\neq 0} \hbar\bigg[ (\omega_{p}l ) b_l^\dagger b_l^{\phantom{\dagger}} 
 + i b_l^\dagger (\Omega_l a_l^{\phantom{\dagger}} + \Omega_{-l} a_{-l}^{\dagger}) + \mathrm{h.c.} )  \bigg]\, + H_{\mathrm{cav}},} + V_{\mathrm{anh}}$} \label{eq:edgeHamiltonian}
\end{align}



\noindent with the light-matter coupling strength $\Omega_l = e/h E_l L\sqrt{\nu/|l|} = 2\omega_c/(\sqrt{|l|}\omega_{\mathrm{cyc}}) \Omega_{\mathrm{Rabi}} \propto \sqrt{N_e}$, where $\Omega_{\mathrm{Rabi}} = eA_l \sqrt{N_e\hbar \omega_{\mathrm{cyc}}/(2m_e\hbar^2)}$ is the collective Rabi frequency~\cite{ciuti_cavity-mediated_2021}. This implies that an absorbed photon, with angular momentum $l$, creates an edge plasmon with linear momentum $k_l= 2\pi l/L$, see Figs.~\ref{fig:figure2}(b)-(d). Crucially, the total angular momentum is thus conserved, as the Hall system can absorb both SAM and OAM. $V_{\mathrm{anh}}$ incorporates anharmonic corrections to the confinement potential, coupling the $l=1$ center-of-motion plasmon mode with higher multipoles ($l>1$), thereby violating Kohn's theorem. However, these corrections scale at least as $\mathcal{O}(1/N)$, making them negligible in the large-$N$ limit~\cite{nardin_linear_2023, wen_theory_1992}.     The following points are important to note: (i) In the case of a cavity field with opposite chirality $l<0$ to that of the edge mode, we obtain a counterrotating coupling term $i\Omega_{-l}( b_l a_{-l} - b_l^\dagger a_{-l}^\dagger)$ in Eq.~\eqref{eq:edgeHamiltonian}, indicating that an absorbed photon destroys a plasmon; (ii) 
 In small systems, noncircular terms that mix different angular momenta appear in the Hamiltonian; these can be treated perturbativly within our theory~\cite{fern_effective_2018}; 
  (iii) Geometry: in the main text we focus on a disk, however a similar diagonal Hamiltonians also apply to other geometries (Hall-bar, Corbino), with geometry only entering via form factors (see supplemental material)  
(iv) Including dipole self-energy slightly renormalizes the plasmon frequency and light-matter coupling~\cite{supmat}; and (v)
we can derive the same photon-plasmon coupling through a cavity-induced anomalous current~\cite{futureAnomaly}.

Strong light-matter coupling typically leads to the formation of polaritons~\cite{carusotto2013quantum}. For simplicity, we first consider a single-mode cavity with an angular momentum $l$.  This limit can be realized using chiral cavities, which are a topic of active experimental research~\cite{hubener_engineering_2021}.  Due conservation of angular momentum, the cavity will couple solely to a single plasmon mode. We can diagonalize the Hamiltonian~\eqref{eq:edgeHamiltonian} via a Bogoliubov transform and obtain the polariton eigenenergies $2E_\pm = \hbar\omega_p l + \hbar \omega_c \pm \sqrt{\Delta^2 + 4\hbar^2 \Omega_l^2}$, where $\Delta = \hbar(\omega_c - \omega_p l)$. The resulting spectrum reveals typical anti-crossing, see Fig.~\ref{fig:figure2}(d). Importantly, the anticrossing can be measured using spectroscopy experiments: using common experimental values~\cite{camino_transport_2006, lin_quantized_2021, roosli_characterization_2021}, the edge drift velocity at \(\nu=1/3\) lies in \(v \approx (2\!-\!4)\times10^{4}\,\mathrm{m/s}\). For an edge circumference \(L=100\,\mu\mathrm{m}\), the fundamental edge plasmon has \(k=2\pi/L\) and wavelength \(\lambda_p\simeq L\), giving \(f_p=v/L\approx 0.2\!-\!0.4\,\mathrm{GHz}\) \((\omega_p=2\pi f_p)\), well below the cyclotron frequency and the free-space light wavelength is \(\lambda_{\text{light}}\approx0.3\,\mathrm{m}\). Note that in the presence of polariton excitations, the Laughlin state $\ket{\Psi_0}$ is no longer the ground state, potentially leading to modified transport properties.

\begin{figure}[t!]
    \centering
    \includegraphics[width = \columnwidth]{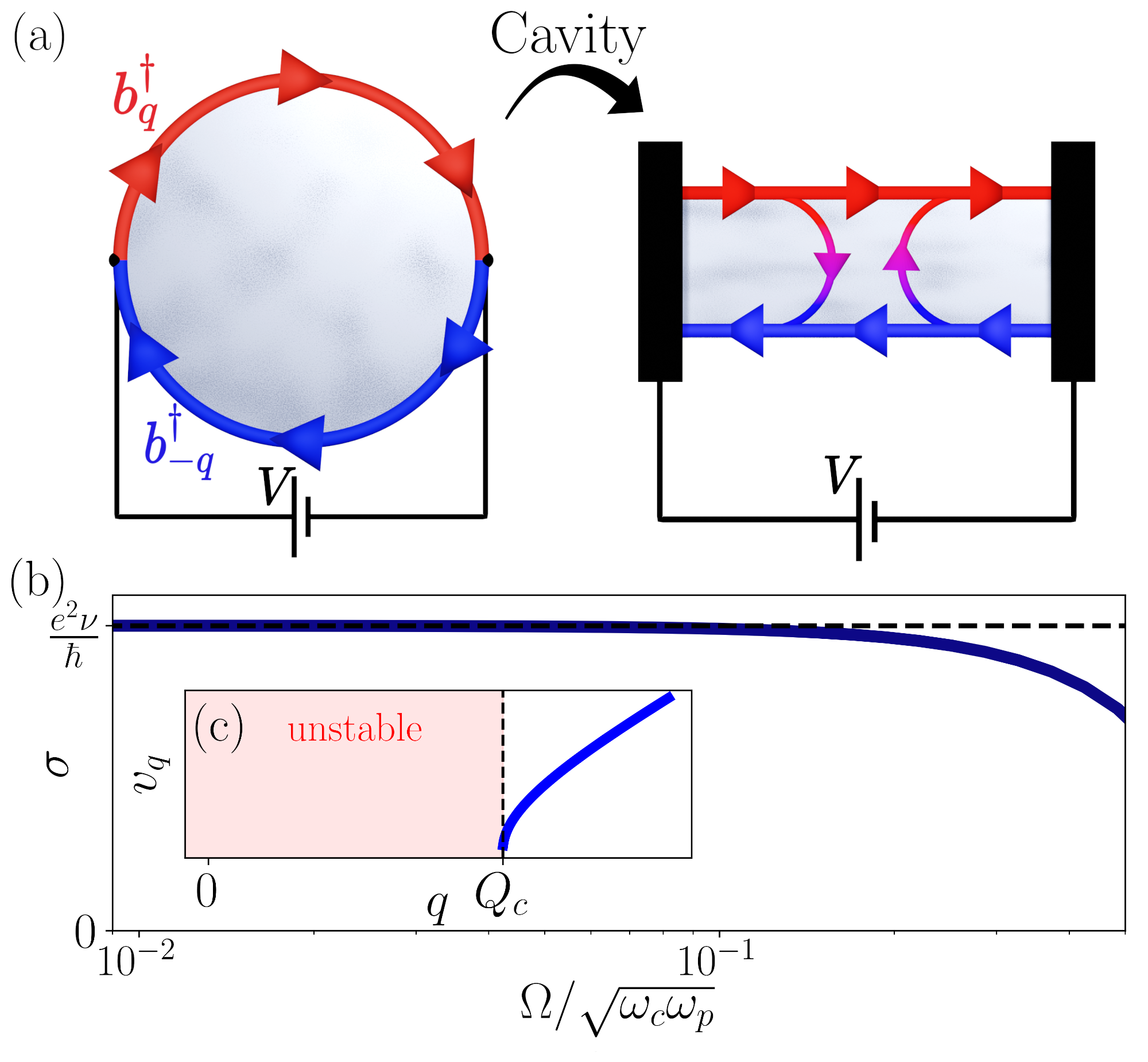}
    \caption{ \textit{Cavity mediated breakdown of topological protection.} (a) The plasmons around the disc can be decomposed into right-moving   (red) and left-moving (blue) modes [cf.~Eq.~\eqref{eq:twoEdgeHamiltonian}]. Cavity-mediated interactions couple the right- and left-movers and transform the system into an effective 1D wire [cf.~Eq.~\eqref{eq:HamEffectiveEdge}]. (b) The Hall conductivity as a function of the normalized coupling strength [cf.~Eq.~\eqref{eq:conductivityMultimode}], 
    (c) Backscattering red-detunes the plasmon dispersion, which can make the plasmon polaritons unstable below the wavevector $Q_c = \frac{4\Omega^2}{\omega_c v}$. }
    \label{fig:figure3}
\end{figure}

We next assess how coupling to the cavity mode affects conductivity, as measured at opposing points on the Hall disc.  As Kohn's theorem is broken, the Hall plateaus may (but do not have to) be disrupted.  The Ohmic contacts effectively partition the edge into two branches,  one with left-moving ($q < 0$) and one with right-moving plasmonic excitations ($q > 0$), see Fig.~\ref{fig:figure3}(a). Inverting the y-axis ($\mathcal{T}: y \to -y$) morphs the left-moving edge into the right-moving edge. Correspondingly, applying $\mathcal T$ to the light-matter coupling Hamiltonian  \eqref{eq:edgeHamiltonian}, yields the coupling to the left-moving edge. Additionally, applying the inversion to the plasmon wavefunction coordinates leads to complex conjugation $\mathcal{T} z = z^*$, so $\mathcal T b_q = b_{-q}^\dagger$. Together, the  Hamiltonian containing both edges reads
\begin{align}
    \resizebox{0.9\hsize}{!}{$\displaystyle{H_{2E} = \sum_{q\neq 0} \hbar\bigg[ v|q| \  b_q^\dagger b_q^{\phantom{\dagger}} + 
 i b_q^\dagger (\Omega_q a_{l_q}^{\phantom{\dagger}} + \Omega_{-q} a_{-l_q}^{\dagger}) + \mathrm{h.c.}   \bigg]\, + H_{\mathrm{cav}}  \, ,}$}\label{eq:twoEdgeHamiltonian}
\end{align}
\noindent Note, we neglect anharmonic corrections to the confinement potential. However, as these corrections do not affect universal properties like conductivity and diminish with system size ~\cite{nardin_linear_2023, wen_theory_1992}, we expect  our results to hold across the entire Hall universality class.   To ascertain the effective cavity-mediated coupling between the left- and right-moving edge states, we transform Eq.~\eqref{eq:twoEdgeHamiltonian} using a Schrieffer-Wolff transformation
\begin{align}
\resizebox{0.9\hsize}{!}{$\displaystyle{H_{\mathrm{eff}}= \sum_{q\neq 0} \hbar|q|\bigg[  \bigg(v + V(q)\bigg)b_{q}^\dagger b_q^{\phantom{\dagger}} -   \frac 1 2 V(q)(b_q^\dagger b_{-q}^\dagger + b_{q}^{\phantom{\dagger}} b_{-q}^{\phantom{\dagger}})\bigg] \, .}$}\label{eq:HamEffectiveEdge}
\end{align}
\noindent Here, $V(q) = -2\Omega^2\omega_c/(|q|(\omega_c^2 - v^2q^2))$ with $\Omega = \Omega_1$~\cite{supmat}. Importantly, we obtain the familiar Luttinger Hamiltonian for a 1D wire when the cavity is included already within second-order perturbation theory~\cite{giuliani_quantum_2005}. Specifically, the cavity introduces plasmon backscattering proportional to $V$, see Fig.~\ref{fig:figure3}(a). This is a key result of our work.

Using our effective (matter-only) model, we can compute the conductivity through the opposite edges of the device, see Fig.~\ref{fig:setup}(a). To this end, we diagonalize the effective model~\eqref{eq:HamEffectiveEdge} using a Bogoliubov transformation $b_q^\dagger = \cosh(\varphi_q)\beta_q^\dagger - \sinh( \varphi_{q}) \beta_{-q}$, where $\tanh(2\varphi_q) = V/(v + V)$, and $b_q$ transforms accordingly~\cite{giuliani_quantum_2005}. In diagonal form, the model reads $H_{\mathrm{eff}} = \sum_{q \neq 0} \hbar v_q |q| \beta_q^\dagger \beta_q^{\phantom{\dagger}}$, with the plasmon dispersion $v_q = \sqrt{v^2 + 2vV}$. The current operator is defined by $ j_q = (i/q)\partial_t n_q$, with $n_q$ the electron density. In our case, the current operator linearized around small wavevectors reads $j_q =\sqrt{L|q|\nu/(2\pi)}v\  \mathrm{sgn}(q)(\beta_q - \beta_{-q}^{\dagger})$~\cite{von_delft_bosonization_1998}. Using Kubo's formula for small wavevectors, the conductivity then reads~\cite{bruus_many-body_2004}
\begin{align}
    \sigma = \frac{  e^2 }{\hbar L}\lim_{\omega \to 0}\int_{-\infty}^{0} dx'\sum_{q\neq 0} e^{iq(x-x')} G_{j}(q, \omega)\,, \label{eq:conductivity}
\end{align}
\noindent where $G_j(q,\omega) = i\int_0^\infty dt \ \braket{[j_q(t), n_{-q}(0)]} e^{i(\omega + i\eta)t}$ is the retarded current-density correlation function, where we used the relation $\beta_q^\dagger(t) = e^{iv_q |q| t}\beta(0)$~\cite{giuliani_quantum_2005}. 


We first consider a single-mode cavity with angular momentum $\pm l$, similar to that employed in a recent experiment~\cite{appugliese_breakdown_2022}~\cite{supmat}
\begin{align}
\resizebox{0.9\hsize}{!}{$\displaystyle{ 
    \sigma = \frac{e^2\nu}{h} + \frac {e^2 }{\hbar L}  \sum_{q = \pm 2\pi/L l}\frac i {q} e^{iq(x-x')}\lim_{\omega \to 0}(\tilde G_j - G_j)\nonumber=\frac{e^2 \nu}{h}\, ,} $}
\end{align}
\noindent where $\tilde G_j$ is the current-density correlation function in the presence of the cavity and $G_j$, is the one without it.  The correlation function for individual plasmons vanishes in the DC limit $\lim_{\omega \to 0} \tilde G_j(\pm q_0, \omega) =0$  if transport is measured a the ends of the edge $x-x'=L$. Crucially, for a homogeneous cavity mode $l = 1$ our result is  consistent with Kohn's theorem and complies with recent transport experiments~\cite{appugliese_breakdown_2022}, as the conductivity is not modified by a homogeneous cavity mode, despite coupling to the edge excitations. Note that we consider a fully coherent (Hamiltonian) system, where no current can be dissipated; a dissipative cavity or edge disorder may lead to a weaking of the topological protection~\cite{ciuti_cavity-mediated_2021, rokaj_topological_2023}.


Next, we consider a multimode cavity with highly inhomogeneous mode profiles, disrupting Kohn's theorem. Specifically, we consider a ``pizza-slice aperture'' by reducing the azimuthal angle $\Delta \phi$~\cite{franke-arnold_uncertainty_2004}. This increases the uncertainty in the angular momentum $\Delta L$, e.g.,  as $\Delta L \cdot \Delta \phi \geq  \hbar/ 2$ due to the uncertainty relation, and thus all photon modes participate equally $A_l \to 1$ in Eq. (2). 
Given the continuum of modes, we can replace the sum over $q$ in Eq.~\eqref{eq:conductivity} with an integral, and obtain using the residue theorem ~\cite{supmat}
\begin{align}
    \sigma = \frac{e^2\nu}{h}\sqrt{ 1 - \frac{4\Omega^2}{\omega_c\omega_p}} \ . \label{eq:conductivityMultimode}
\end{align}
 In this inhomogeneous limit, as the light-matter coupling strength $\Omega$ increases, the conductivity deviates from the universal value $e^2\nu/h$ leading to the destruction of conductivity quantization, see Fig.~\ref{fig:figure3}(b). 
 The cavity-induced backscattering
lifts the chiral protection of the edge modes, thus allowing charges on the opposite edges of the 2D bulk to screen each other. 
Specifically, the cavity modifies the plasmon dispersion to $v_{ q} = v\sqrt{1 - 4\Omega^2/(|q|\omega_c v)}$. Furthermore, plasmons with wavevector below $|q| \leq Q_c =   4\Omega^2/(\omega_c v)$ red-shift and can become unstable [see Fig.~\ref{fig:figure3}(c)], signaling a potential phase transition in the system. What phase appears beyond the instability is beyond the scope of our model.  Please note this result specifically applies to multimode cavities and is therefore fully consistent with recent experiments indicating that Hall plateaus are enhanced for a single homogeneous cavity mode~\cite{enkner_enhanced_2024}. 

Three key implications of our result involve (i) the cavity modes can form polaritons with the plasmonic edges of the FQHE, (ii) a homogeneous mode is insufficient for impacting the topological protection, 
and (iii) an inhomogeneous multi mode cavity  can destroy the Hall plateaus. 
The latter is a result of light-induced backscattering between opposite edges of the Hall system. Our model provides a versatile platform for studying light-matter interaction in the FQHE, which can be readily extended to other filling fractions $\nu = 1/(2p + 1)$, with multiple edge modes~\cite{wen_theory_1992}. We anticipate similar coupling between light and topological boundary modes in other systems that exhibit the chiral anomaly. Taking into account realistic experimental values, our predictions are within reach of experimental observation, and motivate spectroscopic methods for probing the topological order of the FQHE, as well as assist in resolving long-standing controversies concerning the system’s edge structure~\cite{johnson_composite_1991, wen_topological_1995}. Interestingly, the decay of the edge plasmons into photons opens up the potential of controlling the orbital angular momentum of light. Beyond mesoscopics, our model can be applied cold atomic systems, as well as other quantum simulators~\cite{leonard_realization_2022,kumar_cavity_2021}.

\begin{acknowledgments}
We thank S.~Simon, P.~Rabl, C.~Ciuti, and I.~Carusotto for fruitful discussions. The authors acknowledge financial support from the Deutsche Forschungsgemeinschaft (DFG) - project number 449653034 and SFB1432.
\end{acknowledgments}

\nocite{di_francesco_conformal_1997}
\nocite{tong_lectures_2016}

\bibliography{Citations}

\end{document}


\onecolumngrid
\suppmaterial{\mytitle}{\myauthor}{\myaffiliation}{\mydate}

\section{Summary of previous work on Kohn's theorem}

In this supplemental material, we summarize previous research on Kohn's theorem in particular in relationship to light-matter interaction in the fractional quantum Hall effect. For further information see Refs. \cite{kohn_cyclotron_1961, k_jain_jainendra_composite_2007, maag_coherent_2016,appugliese_breakdown_2022, rokaj_topological_2023}. We highlight that the coupling of light to the so-called Kohn mode has been recently explored in Ref.~\cite{rokaj_topological_2023}. The Kohn mode refers to the center-of-motion mode of an electron gas.

 Kohn's theorem states that homogenous light fields only couple to the center of motion (COM) degrees of freedom, and therefore the cyclotron resonance frequency is unaffected by interactions~\cite{kohn_cyclotron_1961}. Let us consider Hamiltonian (1) of the main text in the absence of the confinement potential \cite{k_jain_jainendra_composite_2007}

\begin{align}
    H_C =  \frac 1{2m_e}\sum_j (\boldsymbol \pi_j^2 - e\boldsymbol{\pi}_j\boldsymbol \cdot \boldsymbol A(\boldsymbol{r}_j)) + \sum_{i>j}V_{\mathrm{int}}(|\boldsymbol{r}_i -\boldsymbol{r}_j|)  + H_{\mathrm{cav}}\, .  \label{eq:HC} 
\end{align}
We can define the center of motion coordinate of the electrons by $\boldsymbol{R} = \sum_j \boldsymbol{r}_j$. In the absence of a confinement potential, the center of motion (COM) movement of the electrons in the FQHE can be decoupled from the relative degrees of freedom.  
Crucially, if the vector potential is homogeneous $\boldsymbol{A}(\boldsymbol{r}) =\boldsymbol{A}$ it only couples to the center of motion $H_{\mathrm{int}} = \boldsymbol{A}\cdot \sum_j \boldsymbol{\pi}_j = \boldsymbol\Pi \cdot \boldsymbol{A}$, where $\boldsymbol{\Pi} = \sum_j \boldsymbol{\pi}_j$ but not to the relative degrees of freedom. As a result, the Hamiltonian~\eqref{eq:HC} can be written as \cite{k_jain_jainendra_composite_2007, rokaj_topological_2023}

\begin{align}
    H = \frac 1 {2m} (\boldsymbol{\Pi} - e\boldsymbol{A}) + H_{\mathrm{rel}} + H_{\mathrm{cav}} \ , 
\end{align}

\noindent where $H_{\mathrm{rel}}$ depends on the relative coordinates $\boldsymbol{r}_{ij} = \boldsymbol{r}_i - \boldsymbol{r}_j$ via the interaction potential $V_{\mathrm{int}}$. The coupling to light can therefore be understood in an effective single-particle picture and is not affected by correlations. The collective excitations appear due to the interaction term, i.e., through modifications in the relative degrees of freedom, and therefore do not couple to light. This explains why direct coupling to light was not observed for the collective excitations of the FQHE~\cite{rokaj_topological_2023, appugliese_breakdown_2022}. The coupling of light to the COM in the quantum Hall effect was recently investigated in Ref.~\cite{rokaj_topological_2023}. 

Kohn's theorem can be broken in two ways. First, a non-quadratic confinement potential $V(\boldsymbol r_j) = V(\boldsymbol{R}/2 + \sum_{i\neq j} \boldsymbol r_{ij})$ will mix the COM degree's of freedom with the relative degrees of freedom. This means that the cyclotron resonance frequency will be affected by the interactions. This explains how the light-matter coupling in the main text for the homogeneous $l = 1$ mode manifests. Second, an inhomogeneous vector potential $\boldsymbol{A}(\boldsymbol{r})$ would lead to light that can directly couple to the relative degrees of freedom. In the main text, this is the case for modes with $l>1$.

\section{Short recap on bulk-edge correspondence in the FQHE for Laughlin fractions}

\label{sec:bulkedge}

In this section, we provide a relevant pedagogical overview of the bulk-edge correspondence in the fractional quantum Hall effect (FQHE) for Laughlin fractions based on Refs.~\cite{wen_chiral_1990, fern_structure_2018, simon_wavefunctionology_2020, dubail_edge-state_2012}. First, we introduce the bulk-edge correspondence using qualitative arguments in Sec.~\ref{sec:bulkEdgeIntro}. Then, in Sec.~\ref{sec:PlasmonWavefunc}, we provide an overview on how to explicitly derive the plasmon wavefunctions from the chiral Luttinger liquid ($\chi$LL) edge theory.

\subsection{Pedagogical introduction to the bulk-edge correspondence} 
\setcounter{equation}{0}
\label{sec:bulkEdgeIntro}

As discussed in the main text, the fractional quantum Hall droplet hosts plasmonic excitations on the edge of the system.
When a plasmonic excitation deforms the edge of the FQHE droplet, it affects the behavior of the electrons in the bulk due to the droplet's incompressibility and conservation of current~\cite{fern_structure_2018, wen_chiral_1990}. As a result, it is reasonable to assume that each edge excitation is associated with a bulk wavefunction, which is expected to be similar to the Laughlin wavefunction. In Section~\ref{sec:PlasmonWavefunc}, we will explicitly derive the form of this wavefunction using arguments from conformal field theory (CFT). In this section, we will provide some qualitative arguments what these wavefunctions should look like.

There are three conditions that potential bulk representations $\Psi_l$ of a plasmonic edge excitation $b_l^\dagger$ should satisfy~\cite{fern_structure_2018}

\begin{enumerate}
    \item The edge excitations must be gapless in the absence of a confinement potential. 

    \item The edge excitation $b_l^\dagger$ should possess $m$ quanta of angular momentum more than the Laughlin ground state.

    \item The degeneracy in the spectrum of the edge excitations should be the same as that of the bulk wavefunctions $\Psi_l$.
\end{enumerate}

\noindent Multiplying the Laughlin wavefunction $\Psi_0$ with any symmetric polynomial 
$s(z_1, \dots z_n)$ of the electron positions $z_i$ increases the size of the droplet but keeps the energy at zero. Therefore, the wavefunction must have the general form $\Psi_l= s(\{z_j\})\Psi_0$ to fulfill the first condition. To satisfy the second condition that the wavefunction possesses the appropriate angular momentum, we can choose the symmetric polynomial $s(\{z_j\})= p_l = \sum_k z_j^l$.  Multiplying with a symmetric polynomial $p_l$ increases the total angular momentum $m_0$ of the state by $m$ units $m_{\mathrm{tot}} = m_0 + m$~\cite{wen_theory_1992}. As a result, the total wavefunction is given by~\cite{simon_wavefunctionology_2020}
\begin{align}
    \Psi_l = p_l \cdot \Psi_0 = \sum_i z_i^l  \cdot \Psi_0= \bigg(\sum_i z_i^l\bigg) \cdot \prod_{j<k} (z_j -z_k)^{1/\nu} \prod_k e^{-\frac 1 4 |z_k|^2}\, , \label{eq:edgewave}
\end{align}
where we inserted the definition of the Laughlin ground state for the filling fraction $1/\nu$.
We now need to check if the wavefunctions~\eqref{eq:edgewave} have the same degeneracy structure as the edge excitations, i.e., that they fulfill the third condition. In this case, it makes sense to assume a quadratic confinement potential that gives the energy $\hbar \omega_p m$ to the edge modes according to their angular momentum $m$. This means that multiplying the Laughlin wavefunction with a symmetric polynomial no longer keeps the energy at zero. To obtain a state with energy $\hbar \omega_p$, we can act with $b_1^\dagger$ on the edge vacuum state or multiply the Laughlin wavefunction with $p_1$. This implies that the degeneracy of this first excited state in both the bulk and edge Hilbert spaces is reduced to one. For reaching the $m_{\mathrm{tot}} = 2$ state, we can act with ${b_1^\dagger}^2,\ {b_1^\dagger}$ or multiply the wavefunction with $p_2$ or $p_1^2$, implying that the degeneracy is two~\cite{wen_chiral_1990, simon_wavefunctionology_2020}. The general degeneracy structure is summarized in table~\ref{tab:degeneracy}.

\begin{table}[H]
\begin{tabular}{c | c c c c c c c}
\rowcolor{light_gray} \textbf{$m$} &  1 & 2 & 3  & ...  \\
\hline
{edge} & ${b_1^\dagger}$ & ${b_1^\dagger}^2,\  {b_2^\dagger}$ & ${b_1^\dagger}^3 {b_1^\dagger} {b_2^\dagger},\  {b_1^\dagger}$   & ...  \\
 $\Updownarrow$ &  $\Updownarrow$ & $\Updownarrow$  & $\Updownarrow$  &     \\
{bulk} & $p_1$ & $p_1^2,\  p_2$ & $p_1^3,\  p_1 p_2,\  p_3$  &...  \\
\hline
\rowcolor{light_gray} {degeneracy} & 1 & 2 & 3  & ... \\
\end{tabular}
\centering
\caption{The edge excitation creation operators and bulk symmetric polynomials combinations necessary to generate a state with a total angular momentum of $m$. }
\label{tab:degeneracy}
\end{table}

\noindent The degeneracy structure of the bulk wavefunction is identical to the degeneracy structure of the edge surface waves.  There is no analytical proof that the edge states generate the entire Kac-Moody algebra, but the statement has been checked numerically to high precision~\cite{dubail_edge-state_2012, wen_theory_1992}. 
We can conclude that the bulk wavefunctions $p_l \cdot \Psi_0$ have similar properties as the edge excitations $b_l^\dagger$. This relationship between bulk wavefunctions and correlators on the edge is formalized in Chern-Simons conformal field theory (CFT)~\cite{frohlich1993gauge,hansson_quantum_2017,frohlich2023gauge}. 

\subsection{Deriving the plasmon wavefunctions using the bulk-edge correspondence}

\label{sec:PlasmonWavefunc}
We will now present a comprehensive summary of the CFT methods used to derive the the bulk-edge correspondence following Refs.~\cite{frohlich1993gauge,hansson_quantum_2017, dubail_edge-state_2012, fern_structure_2018, fern_structure_2018,frohlich2023gauge}.
Unlike the Composite Fermion theory, the Chern-Simons theories do not provide a microscopic explanation of the FQHE in terms of probability distributions for individual electrons. Instead, by applying general arguments based on charge conservation and allowed terms in the low-energy limit Lagrangian, CFT identifies a small set of degrees of freedom that can emerge from the complex interactions of many electrons. These effective degrees of freedom describe the low-energy dynamics and long-range order of the FQHE state. Thus, CFT is well-suited for formulating the topological properties of FQHE states in a clear and concise manner~\cite{fern_structure_2018}. While CFT is a rich mathematical topic, for this text it is sufficient to see CFT simply as a collection of helpful operators for building FQHE wavefunctions. 

The bulk-edge correspondence is a direct result of the quantization of Chern-Simons theories. The Chern-Simons theory describing the quantum hall state is essentially $3 = 2 + 1$ dimensional. 
 If we cut the CFT in a ``time-like'' ($1 + 1$ dimensional) slice, we acquire the CFT of the quantum hall edge. If we cut the CFT at a fixed time in a ``space-like'' ($2 + 0$ dimensional) slice, we obtain a CFT describing the ground state. This fact is called the bulk-edge correspondence~\cite{hansson_conformal_2007}.

First, we recall some fundamental definitions in the $\chi$LL theory of the fractional quantum Hall effect. One important quantity is the edge electron field operator $\Psi(x)$ destroying one electron at the position $x$ on the edge. 
The electron field operator $\Psi(x)$ can be expressed using the bosonic field $\phi(x)$ representing the edge plasmons ~\cite{von_delft_bosonization_1998}.
\begin{align}
    \Psi(x) \propto e^{-i\frac 1 {\nu} \phi(x)}\, , \label{eq:bosonization}
\end{align}
where $\phi(x)$ is related to the edge electron density via $\rho(x) = \frac 1 {2\pi}\partial_x \phi(x)$, and can be written using the plasmonic operators on the edge as 
\begin{align}
    \phi(x) = -\sqrt{\frac{2\pi \nu}{L}}\sum_{k>0} \frac 1 {\sqrt k}\bigg(e^{-ikx} b_k^\dagger + e^{ikx} b_k\bigg)e^{-ak/2}\, ,\label{eq:phix}
\end{align}
where $k \in \mathbb N$  $L$ is the edge circumference, and $a$ is the cutoff of the theory. 

\begin{figure}
    \centering
    \includegraphics[width = 0.4\textwidth]{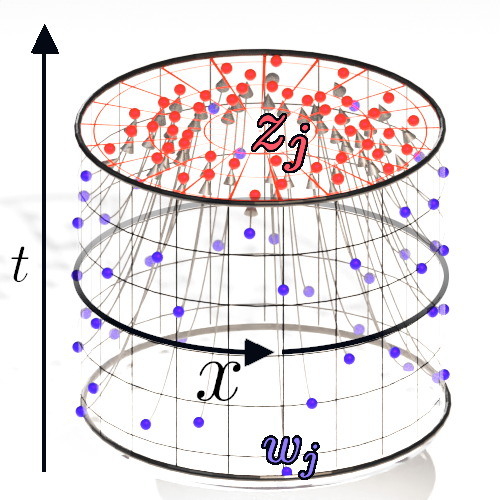}
    \caption{\textit{The state-operator map.} The surface of the cylinder corresponds to the $1+1$-dimensional slice of the Chern-Simons CFT. The surface of the cylinder is parametrized by $x$, the position on the edge of the Hall droplet, and time $t$. The disk inside the cylinder represents the $2+0$-dimensional bulk CFT. The disk is parametrized by the complex variable $z$ representing the complex electron position in space. By utilizing the state-operator map, a blue point located at $w_j = \frac{2\pi}{L} x + it$ is projected onto the red disk representing the bulk at $z_j = x_j + iy_j$. Points at the beginning of time, $t = -\infty$, are mapped to the center of the disk.
 }
    \label{fig:confMap}
\end{figure}

We illustrate the inner working of constructing wavefunctions using CFT by constructing the edge wavefunctions~\eqref{eq:edgewave} introduced in Sec.~\ref{sec:bulkedge} using qualitative arguments. To obtain the wavefunctions, we need to apply a rotation to the time dimension to map the $1+1$ dimensional Lorentzian slice to a $2+0$ dimensional Euclidean space. The mapping can be achieved by Wick rotating the time dimension~\cite{simon_wavefunctionology_2020}
\begin{align}
    w = \frac{2\pi}{L}x_{\mathrm{edge}} + it\ , \qquad w^* = \frac{2\pi}{L}x - it\, . \label{eq:ConfMapping}
\end{align}
We can now complete the mapping by considering the complex electron position $z = e^{-iw} = e^{t}e^{-i\frac{2\pi}{L}x_{\mathrm{edge}}}$. 
We can imagine this mapping in the following way. The time-like slice of the CFT (edge) lives on a cylinder. The circumference of the cylinder, represents the edge at equal time and the vertical direction is the time dimension; see Fig.~\ref{fig:confMap}. The bulk is represented by a disc at equal time $t$. The mapping~\eqref{eq:ConfMapping} now maps a point from the edge of the cylinder to the bulk. A point at $t = -\infty$ is mapped to the center of the disc~\cite{fern_structure_2018}. 

Applying the Wick rotation~\eqref{eq:ConfMapping} to the definition of the bosonic field~\eqref{eq:phix} yields
\begin{align}
    \phi(z) =\phi_0 -i b_0 \ln(z)  - \sqrt\nu \sum_{l>0} \frac 1 {\sqrt l}\bigg(b_l^\dagger z^{-l}  +  b_l z^l\bigg)\, .\label{eq:phiz}
\end{align}
Here, we have introduced the zero mode of the boson field $\phi_0$ and $b_0$ that maintain $[\phi_0, b_0] = i$~\cite{fern_structure_2018}. We have neglected the zero mode when considering the edge, because it decouples from the edge dynamics and is therefore irrelevant~\cite{fern_structure_2018}. Using Eq.~\eqref{eq:phiz} it is possible to show that $\braket{\phi(z)\phi(w)} = -\ln(z-w)$. 
We wick rotate the electron field operator on the edge $\Psi(x,t)$ [cf.~Eq.~\eqref{eq:bosonization}], giving rise to the electron field operator of the bulk $\Psi(z)$
\begin{align}
    \Psi(z) = :e^{i\frac 1 {\nu} \phi(z)}:\, . \label{eq:CFTVertex}
\end{align}
where $:\ \ :$ denotes normal ordering.

For calculating the electron correlator, we need one more aspect. In the plasma analogy of the Laughlin wavefunction, there is a fixed background positive charge density that neutralizes the negative charge of the electrons. This positive charge background is essential for maintaining the charge neutrality of the system and ensuring that the correlator defining the Laughlin wavefunction does not vanish~\cite{hansson_conformal_2007}. The background charge operator is given by~\cite{moore_nonabelions_1991}

\begin{align}
    \mathcal O = \exp\bigg( -\frac i {2\pi l_B^2  }\int dz \ \phi(z) \bigg)\ . \label{eq:laughlinBackCharge}
\end{align}

We can now compute the bulk electron wavefunction by evaluating the CFT correlator over the electron annahilation operators $\Psi(z)$~\eqref{eq:CFTVertex} and the background charge operator $\mathcal O$~\eqref{eq:laughlinBackCharge} by

\begin{align}
    \bigg\langle \prod_j \Psi(z_j) \mathcal O\bigg\rangle = \exp\bigg(-\frac 1 \nu \sum_{j< k} \braket{\phi(z_j) \phi(z_k)}\bigg) \exp\bigg( \sum_k -\frac 1 {2\pi l_B^2 \nu} \int dz \braket{\phi(z_k) \phi(z)}\bigg) \ .\label{eq:laughlinblock}
\end{align}

\noindent Here, we used the fact that the correlator over a free boson field is given~\cite{di_francesco_conformal_1997}

\begin{align}
    \braket{0|:e^{A_1}::e^{A_2}:\dots :e^{A_n}:|0}  = \exp\bigg( \sum_{j<k} \braket{0|A_jA_k|0} \bigg)\, .\label{eq:VertexTrick}
\end{align}

\noindent We can substitute the correlator $\braket{\phi(z_j) \phi(z_k)} = -\ln(z_j - z_k)$ into Eq.~\eqref{eq:laughlinblock}. Regarding the integral in Eq.~\eqref{eq:laughlinblock} the imaginary part can be undone using a gauge transform. The real part results in the Gaussian factor in the Laughlin wavefunction. In total we obtain~\cite{fern_structure_2018}

\begin{align}
    \bigg\langle \prod_j \Psi(z_j) \mathcal O\bigg\rangle = \prod_{j < k} (z_j - z_k)^{1/\nu}e^{-\frac 1 {4l_B^2} \sum_k  |z_k|^2} \, . \label{eq:LaughlinFromCFT}
\end{align}

This shows that the Laughlin wavefunction can be obtained from CFT correlators~\eqref{eq:LaughlinFromCFT}. Consequently, by using the bulk-edge correspondence, we can derive the complete wavefunction in the bulk with just the understanding of how electrons behave on the edge. Consequently, it is possible to achieve a comprehensive topological classification of the bulk by solely relying on the edge.

We can also compute the wavefunctions of different excitations, such as quasiholes or edge plasmons, by inserting the associated operators into the correlator in Eq.~\eqref{eq:laughlinblock}. For example, it is possible to obtain the quasihole wavefunction by inserting the operator $e^{i \sqrt{\nu} \phi(\eta)}$. Similarly, by inserting the operator $b_l$  into the correlator~\eqref{eq:LaughlinFromCFT}, it is possible to obtain the bulk wavefunction of the edge excitations~\cite{dubail_edge-state_2012,fern_structure_2018}. For the derivation, we follow Ref.~\cite{fern_structure_2018}. To obtain the plasmon wavefunction, we need to insert $b_l$  into the correlator~\eqref{eq:LaughlinFromCFT}
\begin{align}
     \Psi_l = \bigg\langle b_l \prod_j :e^{i\frac 1 { \nu} \phi(z_j)}: \mathcal O\bigg\rangle\, . \label{eq:edgeCorrelator}
\end{align}
\noindent If we slide the operator $b_l$ all the way to the right side of the correlator~\eqref{eq:edgeCorrelator} we annihilate the vacuum. During this process, we have to keep in mind that it does not commute with the vertex operators as 
\begin{align}
[b_l, e^{i\frac{1}{\nu}\phi(z_j)}] &= e^{i\frac{1}{\nu}\phi(z_j)} \bigg[b_l, -\sum_{l\neq 0}\frac{1}{\sqrt{\nu |l|}}\left(b_l^\dagger z^{-l}+b_l z^l\right)\bigg]\\ &=  \frac  {1} {\sqrt{\nu }}\frac{z^l}{\sqrt{l}}\cdot e^{i\frac{1}{\nu}\phi(z_j)} \, .\label{eq:CommbmVertex}
\end{align}
\noindent However, this implies that the correlator takes the following form
\begin{align}
     \Psi_l= \bigg\langle b_l \prod_j :e^{i\frac 1 {\sqrt \nu} \phi(z_j)}: \mathcal O\bigg\rangle= \frac 1 {\sqrt{\nu l}}\sum_j z_j^l \cdot  \prod_{j < k} (z_j - z_k)^{1/\nu}e^{-\frac 1 {4l_B^2} \sum_k  |z_k|^2}\, . \label{eq:CFTbmwavefunc}
\end{align}
\noindent This is exactly the same form as Eq.~\eqref{eq:edgewave}. 

In summary, using the bulk-edge correspondance it is possible to construct the bulk wavefunction of the Laughlin states. This is the case because bulk and edge are described by a space-like $2+0$ dimensional and a time-like $1+1$ dimensional slice respectively of the same underlying CFT. Using Wick rotation~\eqref{eq:ConfMapping} one can switch between the different slices~\cite{tong_lectures_2016}. 

We now proceed to evaluate the matrix elements of the light-matter coupling Hamiltonian using the above relation. Our objective, as outlined in the main text, is to compute the matrix elements

\begin{align}
\braket{\Psi_{ \boldsymbol{\lambda}}|H_{\mathrm{int}}|\Psi_{\boldsymbol{\mu}}} &= \int \prod dz_j = i\sum_{l\neq 0}\frac{eE_l}{|l|} \left( \braket{\Psi_{\boldsymbol{\lambda}} |\sum_j z_j^{|l|}| \Psi_{\boldsymbol{\mu}}}\alpha_l - \braket{\Psi_{\boldsymbol{\lambda}}| \sum_j z_j^{|l|}| \Psi_{\boldsymbol{\mu}}}^*\alpha_l^\dagger \right) .
\end{align}

\noindent In general, evaluating these matrix elements requires computing a high-dimensional spatial integral over the Laughlin wavefunctions:

\begin{align}
\braket{\Psi_{ \boldsymbol{\lambda}}|\sum_j z_j^l|\Psi_{\boldsymbol{\mu}}} &= \int \prod dz_j \Psi_{\boldsymbol \lambda}^*({z_j}) \left( \sum_j z_j^l \right) \Psi_{\boldsymbol \mu}({z_j}) .
\end{align}

\noindent However, by leveraging the bulk-boundary correspondence~\cite{fern_effective_2018, dubail_edge-state_2012}, we can simplify the evaluation of these integrals. Specifically, we utilize the relation $\ket{\Psi_{\boldsymbol \mu}} = \prod_j e^{i\phi(z_j)/\sqrt \nu} \ket{\boldsymbol \mu}$. Using the commutator~\eqref{eq:CommbmVertex}, we find that

\begin{align}
\sum_j z_j^l\ket{\Psi_{\boldsymbol \mu}} = \left( \sum_j z_j^l \right) \prod_j e^{i\phi(z_j)/\sqrt{\nu}} \ket{\boldsymbol \mu} = \sqrt{\nu l} \prod_j e^{i\phi(z_j)/\sqrt{\nu}} b_l \ket{\boldsymbol \mu} ,.
\end{align}

\noindent This implies that we can replace

\begin{align}
\braket{\Psi_{ \boldsymbol{\lambda}}|\sum_j z_j^l|\Psi_{\boldsymbol{\mu}}} = \sqrt{\nu l} \braket{\boldsymbol{\lambda}| b_l|\boldsymbol \mu} .
\end{align}

\noindent Thus, we have demonstrated how to evaluate the matrix elements in Equation~(4) of the main text.

\newpage

\section{Gauge transforms to express the light-matter coupling Hamiltonian in terms of Laughlin wavefunctions}
\setcounter{equation}{0}

To compute the matrix elements with the Laughlin wavefunctions in the main text, it is vital to express both the Hamiltonian and the cavity vector potential in an appropriate gauge. This means that we need to write the cavity vector potential as a holomorphic function, as the Laughlin wavefunctions also represent holomorphic functions. We will discuss this procedure in section~\ref{sec:DerivCoupling}.
Next, it is advantageous to express the light-matter coupling term in the Hamiltonian as a function of the electrons' positions instead of momenta. To achieve this, we will apply the dipole gauge discussed in section~\ref{sec:DipoleGauge}.

\subsection{Projecting the vector potential into the lowest Landau Level}
\addtocounter{equation}{1}
\label{sec:DerivCoupling}

\label{sec:cavityCoupling}
    
In the main body, we explored the derivation of the light-matter coupling Hamiltonian specifically for the mode profile described in Eq.~(2) in the main text. However, it is important to note that the angular momentum of light can be used as a basis to describe various mode profiles, such as Laguerre-Gaussian modes, Bessel modes, or hypergeometric Gaussian modes~\cite{padgett_lights_2004, forbes_orbital_2021}. In this section, we demonstrate how to express an arbitrary vector potential such that it lies in the lowest Landau level.

In general, the vector potential is a three-dimensional vector field, represented by 
$\boldsymbol
A(x,y,z)$, that depends on all spatial coordinates. However, since the electrons of the fractional quantum Hall effect can only move in the 2DEG, which is parameterized by 
$(x,y)$, when coupling to them, we must project the vector potential onto their plane. This involves setting 
$z = 0$ and neglecting the $A_z$ component, as this component exerts an out-of-plane force on the electrons. Consequently, the vector potential of a single mode that carries a clearly defined angular momentum of light may typically be expressed as

\begin{align}
    \boldsymbol{A}(\boldsymbol{r}) = A(z) \boldsymbol{\epsilon} + A^*(z) \boldsymbol{\epsilon}^*   \ . \label{eq:totalA}
\end{align}

\noindent The FQH wavefunctions are expressed as holomorphic functions of the electron position $z = x+ iy$. 
To project the light-matter coupling Hamiltonian $H_{\mathrm{int}} = \frac e l \boldsymbol{E} \cdot \boldsymbol{r}$ on the plasmon subspace, it is convenient to express the mode profile $A(z)$ itself as a holomorphic function of the complex variable $z$ . This can be achieved by a gauge transform $\chi$ under which the vector potential $\boldsymbol{A}$ and the scalar potential $V$ transform like 

\begin{align}
    \boldsymbol A &\to \boldsymbol A + \nabla \chi = \boldsymbol{\widetilde A}\ , \\
    V &\to V - \partial_t \chi  = \widetilde V\ ,
\end{align}

\noindent where we take $\chi(x,y)$ as a real valued function. 
 When considering $\boldsymbol{\epsilon} = (1, i, 0)/\sqrt 2$ in the case of circular polarization, we obtain 

\begin{align}
    \boldsymbol{A} = \frac 1 {\sqrt{2}}\begin{pmatrix}
        \Re \{ A(z)\} \\ -\Im \{ A(z)\}
    \end{pmatrix} \to \boldsymbol{A} + \nabla \chi = \frac 1 {\sqrt{2}}\begin{pmatrix}
        \Re \{ A(z) + \partial_x \chi\} \\ -\Im \{ A(z) - i \partial_y \chi \}
    \end{pmatrix}   \ .
\end{align}

\noindent This implies that under a gauge transformation, $A(z)$ transforms as $A(z) \to A(z) + \partial_x \chi - i \partial_y \chi$. Now, let's consider a complex-valued scalar $\chi(z = x + iy)$, where $2d_z \chi =  \partial_x \chi(x,y) - i \partial_y \chi(x,y)$. By introducing a complex $\chi(z)$, we can express the gauge transformation simply as $A(z) = A(z) + d_z \chi(z)$.

Our objective now is to determine a gauge transformation $\chi(z)$ such that $A_{\mathrm{hol}} = A(z) + d_z \chi$ is holomorphic. According to the Cauchy-Riemann conditions, $A_{\mathrm{hol}}$ should not depend on the antiholomorphic variable, i.e., $d_{(z^*)} A(z) = 0$. This implies that $A_{\mathrm{hol}}$ must satisfy the differential equation:
\begin{align}
    \frac{d}{dz^*} \widetilde A&=\frac{\mathrm{d}}{\mathrm{d} z^*} \bigg(A + \frac{d}{dz} \chi\bigg) = 0 \,,\\ 
    \frac{\mathrm{d}^2\chi}{(\mathrm{d} z )(\mathrm{d}z*)} &= - \frac{\mathrm{d} A(z)}{\mathrm{d}z^*}\,.  \label{eq:gaugeDiffEq}
\end{align}
\noindent In turn, this implies that by solving the differential Eq.~\eqref{eq:gaugeDiffEq}, we can determine the suitable gauge transformation that renders the mode profile holomorphic. Equation~\eqref{eq:gaugeDiffEq} bears similarity to the Poisson equation $\Delta \chi = -\nabla\cdot A$, which one must solve to express a vector potential in the Coulomb gauge~\cite{jackson_lorenz_2002}. 

In polar coordinates $z = r e^{i\varphi}$,  Eq.~\eqref{eq:gaugeDiffEq} can be written as 
\begin{align}
    \frac{e^{-i\phi}} 2 \bigg(\frac 1 {r^2}\partial_\phi^2  + \frac 1 r \partial_r  + \partial_r^2\bigg) \chi= - \bigg( \partial_r - \frac 1 {ir} \partial_\phi\bigg) A(r,\varphi)\ .
\end{align}
In the special case that the mode profile $A(z)$ is an eigenstate of the angular momentum operator $L_z = -i\hbar \partial_\varphi$, we can write it as $A(z) = A(r) e^{im\varphi}$. In this case, we can use the Ansatz $\chi(r,\varphi) = \chi(r) e^{il\varphi}$ where $l = m+1$. By solving the differential equation
\begin{align}
    \frac{1} 2 \bigg(\frac {l^2} {r^2} \chi - \frac 1 r \partial_r \chi - \partial_r^2\chi\bigg) =  \bigg( \partial_r + \frac {l-1} {r} \bigg) A(r)\,,
\end{align}
\noindent we can determine the radial profile of the gauge transform $\chi$. 

After obtaining the holomorphic mode profile $\widetilde A(z)$ corresponding to $\boldsymbol{\widetilde A} = \widetilde A(z) \boldsymbol{\epsilon} + \widetilde A(z)^* \boldsymbol{\epsilon}^*$, we can now quantize it in terms of angular momentum modes. As the plasmon wavefunctions are holomorphic, it is reasonable to apply holomorphic quantization. This means that we express the vector potential as a Taylor series in the holomorphic variable $z$
\begin{align}
    A(z) = \sum_{m=  0} A_{0,m} z^m a_m  \, ,
\end{align}
where 
 the Taylor coefficients $A_{0, m}$ are given by $A_{0, m} = \frac{d^m A(z)}{dz^m} \vert_{z = 0}$ and $a_m$ represents a complex amplitude. Using $l = m + 1$ and $A_{0,l} = A_{0, m+1}$, the total vector potential then takes the form 
\begin{align}
    \boldsymbol{\widetilde A} = \sum_{l\neq  0} A_{l} z^{|l| - 1} a_l + A_{l}^* (z^*)^{|l|-1} a_l^\dagger \ , \label{eq:tildeAA}
\end{align}

\noindent with $\alpha_l = \Theta(l) a_l + \Theta(-l) a_l^\dagger$ and $\alpha_l^\dagger = \Theta(l) a_l^\dagger + \Theta(-l) a_l$.
The quantization is complete when promoting $a_l$ and $a_l^\dagger$ to quantum mechanical operators with $[a_l, a_{l'}^\dagger] = \delta_{l,l'}$ and $[a_l, a_{l'}] = [a_l^\dagger, a_{l'}^\dagger] = 0$. Then, Eq.~\eqref{eq:tildeAA} has the same form as the vector potential presented as Eq.~(2) in the main text.

\subsubsection{Example: Hypergeometric Gaussion modes}

In the main text, we have considered a Hypergeometric Gaussian mode profile with a broad beam size $A(z) = z^m = r^m e^{im\phi}$.
In this section, we will consider HG modes, where the beam width $\sigma$ is finite~\cite{kotlyar_hypergeometric_2007,karimi_hypergeometric-gaussian_2007}

\begin{align}
A_{\mathrm{HG}} = A_{0, \mathrm{HG}} \frac{r^m}{\sigma^m}e^{-r^2/(2\sigma^2)}e^{il\phi}\  .  \label{eq:HGMode}
\end{align}

\noindent We see that the beam has a finite width, denoted as $\sigma$, defined by the Gaussian factor $e^{-r^2/(2\sigma^2)}$. It is convenient to introduce the normalized coordinates $\rho = r/\sigma$.

This mode profile is written in the Lorenz gauge, and is obtained as a solution to the wave equation ~\cite{kotlyar_hypergeometric_2007}. Using the Cauchy-Riemann conditions in polar coordinates, we observe that the field profile is not holomorphic, as.

\begin{align}
    \partial_\rho A - \frac 1 {i\rho} \partial_\phi A = -A_{0, \mathrm{HG}} \rho^{m+1} e^{-\rho^2/2}e^{im\phi} \neq 0. 
\end{align}

\noindent To write $A(z)$ as a holomorphic function, we need to find the gauge transform $\chi$ according to equation~\eqref{eq:gaugeDiffEq}. In polar coordinates, the holomorphic derivative can be expressed as 

\begin{align}
    \partial_z \chi = \frac{e^{-i\phi}} 2 \bigg( \partial_\rho +  \frac 1 {i\rho} \partial_\phi\bigg) \chi \ . 
\end{align}

\noindent This means equation~\eqref{eq:gaugeDiffEq} becomes in polar coordinates 

\begin{align}
    \frac{e^{-i\phi}} 2 \bigg(\frac 1 {\rho^2}\partial_\phi^2 \chi + \frac 1 \rho \partial_\rho \chi + \partial_\rho^2\chi\bigg) = - \bigg( \partial_\rho - \frac 1 {i\rho} \partial_\phi\bigg) A.
\end{align}

\noindent To find the appropriate gauge transform for the Hypergeometric Gaussion modes, we solve the following second-order inhomogeneous differential equation 

\begin{align}
    \frac 1 \rho \partial_\rho \chi + \partial_\rho^2\chi  +\frac 1 {\rho^2}\partial_\phi^2 \chi  = 2A_{0, \mathrm{HG}}  \rho^{m+1} e^{-\rho^2/2} e^{i(m+1)\phi} \ . \label{eq:DiffHGMode}
\end{align}

\noindent We will now apply separation of variables by making the Ansatz

\begin{align}
    \chi(\rho, \phi) =  \rho^{2} e^{i(m+1)\phi} \chi_\rho(\rho)\ . \label{eq:Ansatz}
\end{align}

\noindent Using the Ansatz~\eqref{eq:Ansatz} equation~\eqref{eq:DiffHGMode} becomes 

\begin{align}
    \rho^2 \partial_\rho^2 \chi_\rho  + 5\rho \ \partial_\rho \chi_\rho -(m+3)(m-1)\chi_\rho  = -2\rho^{m+1} e^{-\rho^2}. \label{eq:DiffHGModeSimplified}
\end{align}

\noindent We can solve this equation by making a power series Ansatz informed by the right-hand side of equation~\eqref{eq:DiffHGModeSimplified}

\begin{align}
    \chi_\rho(\rho) = \rho^{m+1} \sum_k a_k \rho^{2k}. \label{eq:AnsatzDiffHG}
\end{align}

\noindent Inserting equation~\eqref{eq:AnsatzDiffHG} into equation~\eqref{eq:DiffHGModeSimplified} yields the following equation for $a_k$

\begin{align}
    a_k =A_{0, \mathrm{HG}}  \frac{(-1)^k}{k!} \frac 1 { (2k+m+1)(2k+m+5) - (m+3)(m-1)}. 
\end{align}

\noindent Using Mathematica we can explicitly evaluate the sum in equation~\eqref{eq:AnsatzDiffHG} by using the explicit expression for $a_k$. In this case, the total solution to equation~\eqref{eq:DiffHGMode} becomes 

\begin{align}
    \chi(\rho, \phi) = A_{0, \mathrm{HG}}  \frac {e^{i(m+1)\phi}} {4(m+1)} \bigg(\rho^{2(m+1)} (1 - e^{-\rho^2}) - \Gamma(2+m,0, \rho^2)\bigg)\,, 
\end{align}
where $\Gamma$ is the incomplete Hypergeometric function. This means that the holomorphic mode profile is 

\begin{align}
    A_{\mathrm{HG}, \mathrm{hol}} = A_{\mathrm{HG}} + d_z \chi =A_{0, \mathrm{HG}} z^m
\end{align}

We observe that the holomorphic mode profile of the HG modes is identical to the mode profile considered in the previous sections, albeit with a different prefactor. This similarity arises because the HG modes are also eigenstates of the angular momentum operator $L_z = i \partial_\phi$. However, the adjusted prefactor alters the coupling strength, denoted as $\Omega_l$ in the main text, which is proportional to $A_0$. Additionally, there will be a scalar potential $V = \omega_c \chi$ altering the drift velocity of the plasmons. In conclusion, considering a different mode profile as in the main text will renormalize the parameters of our theory. Nevertheless, the qualitative results remain unchanged.

\subsection{The quantum Hall Hamiltonian in the Dipole gauge}

\label{sec:DipoleGauge}

In the main text, we discussed the application of the dipole gauge to the bulk FQH Hamiltonian $H_c$ [cf.~Eq.~(1) in the main text] in order to accurately consider the cavity's perturbation. In this section, we perform the complete calculation implementing the dipole gauge. 

We start by expanding the Hamiltonian in the Coulomb gauge [cf.~Eq.~(1) in the main text] to obtain
\begin{align}
H_C = \sum_j \bigg[\underbrace{\frac 1 {2m_e} \boldsymbol{\pi}_j^2}_{=H_L} - \underbrace{\frac {e}{2m_e} (\boldsymbol \pi_j \cdot \boldsymbol A(\boldsymbol{r}_j)+ \boldsymbol A(\boldsymbol{r}_j) \cdot \boldsymbol \pi_j)}_{=H_{\mathrm{int}}}+\underbrace{\frac{e^2}{2m_e} \boldsymbol A^2(\boldsymbol{r}_j)}_{H_{\mathrm{Dia}}} \bigg]+ \underbrace{\sum_{l\neq 0}\hbar \omega_{c,l} a_l^\dagger a_l}_{=H_C} + V_{\mathrm{int}} + V_{\mathrm{conf}}\ . \label{eq:HCoulombApp}
\end{align}
\noindent Here, and in the following, we will use the canonical momentum  $\boldsymbol{\pi}_j = \boldsymbol p_j - e\boldsymbol{A}_{\mathrm{ext}}(\boldsymbol{r}_j)$. The term $H_L$ captures all single-electron contributions, $H_{\mathrm{int}}$ denotes the light-matter coupling term in the Coulomb gauge, and $H_C + H_{\mathrm{Dia}}$ denotes the Hamiltonian of the cavity.

To apply the dipole gauge, we utilize a unitary transformation $U$ on the bulk Hamiltonian~\eqref{eq:HCoulombApp}, yielding $H_D = UH_CU^\dagger$. The objective is to express the light-matter coupling term, which involves both photon and electron operators, in terms of the electron position $\boldsymbol r$ rather than the momentum $\boldsymbol{p}$. For the cavity vector potential given in Eq.~(2) in the main text, this transformation can be achieved using the operator
\begin{align}
 U = \prod_j\exp\bigg(\sum_{l\neq 0} \Lambda_l(\boldsymbol{r}_j)\bigg)=\prod_j \exp\bigg( -\frac{i}{\hbar}\sum_{l\neq 0} e \frac{\boldsymbol r_j \cdot \boldsymbol{A}_l(\boldsymbol{r}_j)}{|l|}\bigg) \ , \label{eq:DipoleGauge} 
\end{align}
\noindent where $\Lambda_l = -\frac {i e} {|l|\hbar} \boldsymbol{A}_l \cdot \boldsymbol{r}_j = \frac {i e} {|l|\hbar \sqrt{2}}R  A_{0,l}(  z^l_j \alpha_l + (z_j^*)^l \alpha_l^\dagger)$.

Crucially, both the gauge transform $U(\{\boldsymbol{r}_j\})$~\eqref{eq:DipoleGauge} as well as the interaction potential $V_{\mathrm{int}}(\{\boldsymbol{r}_j\})$ only depend on the position degrees of freedom. This means that the gauge transform commutes with the interaction potential $[V_{\mathrm{int}}, U] = 0$. Therefore, the gauge transform leaves the many-body interaction unchanged $UV_{\mathrm{int}} U^\dagger = V_{\mathrm{int}}$. This implies that

\begin{align}
UH_CU^\dagger = \sum_j U_j \bigg[\frac 1 {2m_e} \boldsymbol{\pi}_j^2 - \frac {e}{2m_e} (\boldsymbol \pi_j \cdot \boldsymbol A(\boldsymbol{r}_j)+ \boldsymbol A(\boldsymbol{r}_j) \cdot \boldsymbol \pi_j)+\frac{e^2}{2m_e} \boldsymbol A^2(\boldsymbol{r}_j)\bigg]U_j^\dagger+ \sum_{l\neq 0} U_l\ \hbar \omega_{c,l} a_l^\dagger a_lU_l^\dagger + V_{\mathrm{int}} + V_{\mathrm{conf}}\ . 
\end{align}

\noindent Here, we defined $U_j = \exp\bigg( -\frac{i}{\hbar}\sum_{l\neq 0} e \frac{\boldsymbol r_j \cdot \boldsymbol{A}_l(\boldsymbol{r}_j)}{|l|}\bigg)$ and $U_l = \exp\bigg( -\frac{i}{\hbar}\sum_{j} e \frac{\boldsymbol r_j \cdot \boldsymbol{A}_l(\boldsymbol{r}_j)}{|l|}\bigg)$. Because the unitary transformation operator $U$ commutes with $V_{\mathrm{int}}$, we can fully understand the effect in terms of a single-particle picture. Hence, in the following, we will suppress the electron index $j$. Please note that the derivation can be performed in the same way in the many-body limit. 

To apply the operator U [cf.~Eq.~\eqref{eq:DipoleGauge}], the Baker-Campbell-Hausdorff lemma is useful: $e^{B} A e^{-B} = A + [B,A] + \frac{1}{2!} [B,[B,A]]+ \dots$. Now, let us compute the fundamental commutators necessary to apply the Baker-Campbell-Hausdorff lemma. The operator $\Lambda_l$ generally commutes with position operators. However, the commutator with the electron canonical momenta $\pi_x$ and $\pi_y$ are not vanishing, as
\begin{align}
    [\Lambda_l, \pi_x] & = eA_{0,l}(z^{|l|-1} \alpha_l + (z^*)^{|l|-1} \alpha_l^\dagger) = e A_x  \ , \label{eq:commgpx}\\ 
    [\Lambda_l, \pi_y] &= eA_{0,l} i(z^{|l|-1}\alpha_l - (z^*)^{|l|-1} \alpha_l^\dagger) = eA_y\ ,\label{eq:commgpy} 
\end{align}
\noindent and $\Lambda_l$ does not commute with the components of the cavity vector potential. Here, we have 
\begin{align}
    [\Lambda_l, A_{x, l'}] &= \frac{eA_{0,l}^2} {|l|\hbar} y|z|^{2(|l|-1)}\delta_{l, l'}\ , \label{eq:commgax} \\
    [\Lambda_l, A_{y, l'}] &= -\frac {eA_{0,l}^2} {|l|\hbar} x|z|^{2(|l|-1)}\delta_{l, l'}\ . \label{eq:commgay} 
\end{align}

\noindent Using the fundamental commutators~\eqref{eq:commgpx},~\eqref{eq:commgpy},~\eqref{eq:commgax}, and~\eqref{eq:commgay}, along with the Baker-Campbell-Hausdorff lemma, we can apply the dipole gauge to the various terms in the Hamiltonian~\eqref{eq:HCoulombApp}. We begin with $H_L$:
\begin{align}
U H_L U^\dagger &= \frac{\boldsymbol \pi^2}{2m_e} + \sum_{l\neq0} [\Lambda_l, \frac{ \boldsymbol \pi^2}{2m_e}] + \sum_{l,l' \neq 0}\frac 1 2[\Lambda_l, [\Lambda_{l'}, \frac{\boldsymbol \pi^2}{2m_e}]]\nonumber\\ 
 &= \frac{\boldsymbol \pi^2}{2m_e} + \frac{e}{2m_e}(\boldsymbol \pi\cdot \boldsymbol A +\boldsymbol A \cdot \boldsymbol \pi)+ \frac{e^2}{2m_e} \boldsymbol A^2  +\frac{e^2\Lambda_0^2}{m_e\hbar} \Upsilon(|z|) (\boldsymbol{\pi} \times \boldsymbol{r})\cdot \boldsymbol{\hat e}_z \ .\label{eq:HLD}
\end{align}
\noindent where we used $A_x = \sum_{l\neq0} A_{x,l}$ and $A_x^2 = \sum_{l\neq 0} A_{x,l}\cdot \sum_{l'\neq 0} A_{x,l'}$. Additionally, we defined 
\begin{align}
    \Upsilon(|z|) = \frac 1 {\Lambda_0^2}\sum_{l\neq 0} \frac {A_{0,l}^2} {|l|} |z|^{2(|l|-1)} = \frac  1 {\Lambda_0^2} \int  d\varphi  |A(z')|^2 , \label{eq:upsilon}  
\end{align}
\noindent  with the normalization constant $\Lambda_0^2 = \int d\varphi |A(z =e^{i\varphi})|^2$. The normalization constant can be thought of as a radially averaged vector potential at the droplet radius.  Please note that $\Upsilon(|z|)$ is a monotonously increasing function with $\Upsilon(0) = 0$ and $\Upsilon(1) = 1$.

Next, we express $H_{\mathrm{int}}$ in the dipole gauge:
\begin{align}
    UH_{\mathrm{int}}U^\dagger &= -\frac{e}{m_e}(\boldsymbol{\pi} \cdot \boldsymbol A + \boldsymbol A\cdot \boldsymbol{\pi}) - \frac{e^2}{m_e} \boldsymbol A^2   
    -\frac{2e^2\Lambda_0^2}{m_e\hbar}  \Upsilon(|z|) (\boldsymbol{\pi} \times \boldsymbol{r})\cdot \boldsymbol{\hat e}_z \ . \label{eq:HintD}
\end{align}

\noindent The diagmagnetic term $H_{\mathrm{Dia}}$ in the the dipole gauge reads
\begin{align}
    U H_{\mathrm{Dia}} U^\dagger &=\frac{e^2}{2m_e} \boldsymbol A^ 2 + \sum_{l\neq 0}[\Lambda_l, \frac{e^2}{2m_e} \boldsymbol A^ 2] +\sum_{l, l' \neq0}\frac 1 2 [\Lambda_{l}, [\Lambda_{l'}, \frac{e^2}{2m_e} \boldsymbol A^ 2]]  \nonumber\\
    &=  \frac {e^2}{2m_e} \boldsymbol A^2 + \frac{e^3\Lambda_0^2}{m_e\hbar } \Upsilon(|z|) (\boldsymbol{A} \times \boldsymbol{r})\cdot \boldsymbol{\hat e}_z + \frac{2e^4\Lambda_0^4}{m_e\hbar^2} \Upsilon^2(|z|)\boldsymbol{r}^2 \ . \label{eq:commDiamag}
 \end{align}
\noindent Last, we calculate the commutator with the cavity $H_c = \sum_{l\neq 0}\hbar \omega_c a^\dagger_l a_l$. Using the form of the vector potential [cf.~Eq.~(2) in the main text], we obtain    
\begin{align}
    UH_C U^\dagger = \sum_{l\neq 0} \frac 1 {|l|} \boldsymbol E_l \cdot \boldsymbol r +\frac{e^2\omega_c \Lambda_0^2}{2\hbar}\Upsilon(|z|) (x^2 + y^2) \ , \label{eq:HCD}
\end{align}
\noindent where we have defined $\boldsymbol E = -i\sum_{l> 0}\omega_{c,l}A_{0,l}(\boldsymbol \epsilon z^l \alpha_l - \boldsymbol \epsilon^*(z^*)^l\alpha_l^\dagger)$.

We can now combine the individual transformed terms~\eqref{eq:HLD},~\eqref{eq:HintD},~\eqref{eq:commDiamag} and~\eqref{eq:HCD}, to obtain
\begin{align}
    H_D = H_L +  \sum_j \bigg[\underbrace{D  \boldsymbol{r}_j^2}_{= H_{\mathrm{conf}}} + \underbrace{\gamma \boldsymbol {\hat e}_z \cdot ( \boldsymbol{\pi}_j \times \boldsymbol{r}_j )  }_{= H_{\mathrm{prec}}}
    \underbrace{+ e \sum_{l\neq 0}\frac 1 {|l|} \ \boldsymbol E_l \cdot \boldsymbol r_j}_{= H_{\mathrm{int, D}}} \underbrace{ -  e\gamma \boldsymbol{\hat e}_z \cdot ( \boldsymbol{r}_j\times  \boldsymbol{A})}_{=H_{\mathrm{int, 2}}}  \bigg] H_C + V_{\mathrm{conf}} + V_{\mathrm{int}}\ , \label{eq:HamDipole}
\end{align}
\noindent where we reintroduced the summation over the particles and $D(|z|) = 1/2\left(e^2\omega_c\Lambda_0^2/\hbar \Upsilon(|z|) + 2e^4/(m\hbar^2)\Lambda_0^4\Upsilon(|z|)^2\right)$, and $\gamma(|z|) = e^2\Lambda_0^2/(m_e\hbar )\Upsilon(|z|)$. 

We can now determine which terms are relevant in the thermodynamic limit.
For this purpose, we compare the energy of each additional term with the energy scale of the light-matter coupling interaction $H_{\mathrm{int}} = e \sum_{l\neq 0}  \boldsymbol{E} _l\cdot \boldsymbol{r}/|l|$.
To this end, we can assume the position is on the order of the magnetic length $|\boldsymbol{r}| \propto l_B = \sqrt{\hbar/(eB)}$
and $|\boldsymbol{\pi}| \propto \hbar/l_B = \sqrt{\hbar e B}$. Please note that the Rabi frequency in the system is given by $\Omega_{\mathrm{Rabi}} = \frac{e}{\hbar}\Lambda_0 \sqrt{(\hbar \omega_{\mathrm{cyc}}N_e)/(2m_e)}$ with the cyclotron frequency $\omega_{\mathrm{cyc}} = (eB)/{m_e}$~\cite{ciuti_cavity-mediated_2021}.

First, we look at the confining potential term $V_{\mathrm{conf}}$ in Eq.~\eqref{eq:HamDipole}. It lifts the angular momentum degeneracy particularly at the edges, and can shift the cyclotron frequency. Its energy compared to the light-matter coupling Hamiltonian $H_{\mathrm{int}}$ is 
\begin{align}
    \frac{|V_{\mathrm{conf}}|}{|H_{\mathrm{int}}|} \leq  \sqrt{\frac 2{N_e}}\frac{\Omega_{\mathrm{Rabi}}}{\omega_{\mathrm{cyc}}} \bigg( 1 + 4\frac{\Omega^2_{\mathrm{Rabi}}}{N_e\omega_{\mathrm{cyc}}\omega_c} \bigg) \ . 
\end{align}
\noindent Here, we approximated $\Upsilon(|z|) \leq \Upsilon(1) = 1$; cf. Eq.~\eqref{eq:upsilon}. We can assume that the collective Rabi frequency $\Omega_{\mathrm{cyc}}$ is at maximum when on the order of the cyclotron frequency $\omega_{\mathrm{cyc}}$, signaling the strong coupling regime. This means $|V_{\mathrm{conf}}|/]H_{\mathrm{int}}| \leq \sqrt{2/ N_e} (1 + 1/N_e)$. This shows in the thermodynamic limit the additional confinement potential has a vanishing effect.  

Next, we look at the term $H_{\mathrm{prec}}$. This is a precession term leading to a shift in the 
cyclotron frequency. Compared with the interaction Hamiltonian, we have
\begin{align}
    \frac{|H_{\mathrm{prec}}|}{|H_{\mathrm{int}}|} \leq \frac{\Omega_{\mathrm{Rabi}}}{\omega_c} \frac 1 {\sqrt{N_e}} \ .
\end{align}

\noindent  Once more, we see that in the thermodynamic limit $N_e \to \infty$, the precession term is not of great significance. Its presence primarily affects the single-electron spectrum, resulting in small quantitative modifications relevant in small systems. In the bulk, it introduces corrections to the cyclotron frequency, while at the edge, it leads to additional corrections to the plasmon drift velocity.

Last, let us examine the second light-matter coupling term. Here, we find
\begin{align}
    \frac{|H_{\mathrm{int, 2}}|}{|H_{\mathrm{int}}|} \leq \frac 1 {N_e} \frac{\Omega_{\mathrm{Rabi}}^2}{\omega_c \omega_{\mathrm{cyc}}}  \ .
\end{align}
\noindent 
Here too, we observe that this term is not relevant in the thermodynamic limit, where $N_e \to \infty$.

In conclusion, in the strong coupling regime, we can express the Hamiltonian in the dipole gauge as follows:
\begin{align}
    H_D =\sum_j \bigg[\frac{\boldsymbol\pi_j^2 }{2m_e}\boldsymbol + \sum_{l\neq 0} \frac{e}{|l|} \ \boldsymbol E_l(\boldsymbol{r}_j) \cdot \boldsymbol r_j\bigg] +  V_{\mathrm{int}} + V_{\mathrm{conf}}+ H_{\mathrm{cav}} \,.  \label{eq:DipoleGaugeStrongCpupling}
\end{align}
\noindent Here we reintroduced the summation over the electrons. This Hamiltonian~\eqref{eq:DipoleGaugeStrongCpupling} is also presented in the main text as Eq.~(3).

\subsection{Effect of the dipole self-energy on light-matter coupling}

In the main text, we have neglected terms from the Hamiltonian $H_D$~\eqref{eq:HamDipole} irrelevant in the theormodynamic limit meaning in mesoscopic systems. In this section, we intend to further analyze the impact of these terms, deriving the full light-matter coupling Hamiltonian. Because of the inhomogeneous mode profile, these Hamiltonian terms depend on the non-linear function $\Upsilon(|z|)$~\eqref{eq:upsilon}. We will approximate $\Upsilon$ close to the edge as $\Upsilon(|z|) \approx \Upsilon(R) = 1$. Furthermore, we need to approximate the normalization factor 

\begin{align}
    \Lambda_0^2 = \sum_{l\neq0}^{l_{\mathrm{max}}} \frac{A_{0,l}^2}{|l|} = 2A_0^2 \sum_{l\neq 0}^{l_{\mathrm{max}}} \frac 1 l \approx 2A_0^2 \ln(l_{\mathrm{max}}) \ ,
\end{align}

\noindent where we assumed (as in the main text) that \(A_{0,l} = A_0\). We now need to determine the cutoff \(l_{\mathrm{max}}\). In general, a mode profile on the edge \(A(\varphi)\) can be expanded into a Fourier series \(A(\varphi) = \sum_{l} A_l e^{i\varphi l}\). This means that  \(l_{\mathrm{max}}\) is determined by the smallest angle \(1/(\Delta\varphi) = \frac{R}{\Delta x}\), where \(x\) is a position on the edge and \(R\) is the disc radius. The disc radius is \(R = l_b \sqrt{2N_e/\nu}\), and the smallest length scale is the magnetic length \(l_B\), i.e., \(l_{\mathrm{max}} = \sqrt{2N_e/\nu}\). Therefore, we have \(\Lambda_0^2 \approx A_0^2 \ln(2N_e/\nu)\).

We can now apply the bulk-edge correspondence to translate the additional terms in Eq.~\eqref{eq:HamDipole} to the Chiral Luttinger liquid theory. First, there is the additional confinement potential \(H_{\mathrm{conf}} = D(|\boldsymbol{z}|) \boldsymbol{r}^2\). The function \(D(|z|)\) depends on the shape of the mode profile and can generally be highly non-linear. However, we are interested in a low-energy effective theory describing small fluctuations around the ground state. Therefore, we can assume \(x^2 + y^2 \approx R^2\) and, therefore, \(\gamma(|R|) = 1\). This means that  \(D(z) \approx \frac{1}{2} (e^2 \omega_c \Lambda_0^2/\hbar + \frac{2e^4\Lambda_0^4}{m\hbar^2})\). Since the confinement potential is quadratic, we can write it in terms of the Luttinger liquid theory. A similar derivation can be found in Ref.~\cite{fern_effective_2018}. As a first step, we need to express the confinement potential in terms of holomorphic variables \(z = x + iy\). We have \(x^2 + y^2 = |z|^2 = zz^*\).

\begin{align}
    H_{\mathrm{conf}} = \sum_j D \boldsymbol{r}_j^2 = D\sum_j |z_j|^2 = D\sum_j z_j z_j^* \ . 
\end{align}

\noindent We now need to project this operator into the lowest Landau level. For this purpose, we can make the replacement $z_j^* \to 2l_b^2 \partial_{z_j}$\cite{k_jain_jainendra_composite_2007}. Hence, we can write 

\begin{align}
    H_{\mathrm{conf}} = 2Dl_B^2 \sum_j \partial_{z_j} z_j \ . \label{eq:HconfProject}
\end{align}

\noindent For formulating this potential within the edge theory, we can use a similar approach as Ref.~\cite{fern_effective_2018}. The action of $H_{\mathrm{conf}}$ on the wavefunction is 

\begin{align}
    H_{\mathrm{conf}} \ket{\Psi} =  2Dl_B^2 \sum_j \bigg[z_j \partial_{z_j} + 1 \bigg]\bigg\langle{\sum_j \Psi(z_j)\mathcal O}\bigg\rangle \ ,
\end{align}

\noindent where $\Psi(z) = e^{i\frac 1 \nu\phi(z)}$ are the vertex operators defined according to~\eqref{eq:CFTVertex} and $\mathcal O$ is the background charge operator~\eqref{eq:laughlinBackCharge}. We can obtain the corresponding operator by commuting $H_{\mathrm{conf}}$ past the vertex operators. For this purpose, we can use the Ward identity~\cite{di_francesco_conformal_1997} 

\begin{align}
    [L_0, \Psi(z)] = \bigg(z\partial_z + \frac 1 {2\nu}\bigg) \Psi(z)\ .
\end{align}

\noindent Here, $L_0$ is the zero mode of the Virasoro algebra~\cite{di_francesco_conformal_1997}. For the purposes here, it is sufficient to know that the $L_0$ mode counts the angular momentum above the ground state. In terms of the mode expansion it is given by~\cite{di_francesco_conformal_1997} 

\begin{align}
    L_0 = \sum_{l > 0} b_l^\dagger b_l + \mathrm{const} \ . 
\end{align}

Since we are only interested in the edge dynamics, we can neglect the impact on the zero mode and therefore assume that $H_{\mathrm{conf}}$ commutes with the background charge~\cite{fern_effective_2018, tong_lectures_2016}. This gives us the following form for the Hamiltonian on the edge

\begin{align}
   \tilde  H_{\mathrm{conf}} =  2Dl_B^2 L_0 + \mathrm{const} = 2Dl_b^2 \sum_{l> 0} b_l^\dagger b_l  \  = \sum_{l > 0} \hbar \bigg( \frac{\Omega^2}{\omega_c} \frac{\ln(2N_e/\nu)}{N_e} + \frac{\Omega^4}{\omega_c^4} \omega_{\mathrm{cyc}} \frac{\ln^2(2N_e/\sqrt{\nu})}{N_e^2}\bigg) b_l^\dagger b_l\,. 
\end{align}

As a result, we find that the effect of the confinement potential \(H_{\mathrm{conf}}\) is to increase the plasmon frequency \(\omega_p \to \frac{\Omega^2}{\omega_c} \frac{\ln(2N_e/\nu)}{N_e} + \frac{\Omega^4}{\omega_c^4} \omega_{\mathrm{cyc}} \frac{\ln^2(2N_e/\sqrt{\nu})}{N_e^2}\). Please note that in the limit of ultrastrong coupling, as we know from the main text, \(\Omega^2 \approx \omega_c \omega_{\mathrm{cyc}}/2\), which means that the plasmon frequency increases by \(\Delta\omega = \omega_{\mathrm{cyc}} (1/N_e + \omega_{\mathrm{cyc}}^2/(N_e^2 \omega_{c}^2))\). This contribution might be relevant in small systems but is completely irrelevant in the thermodynamic limit \(N_e \to \infty\) considered within this work, since \(\ln(N_e)/N_e \to 0\).

Next, we examine the effect of the precession term \(H_{\mathrm{prec}}\) in the Dipole gauge Hamiltonian \(\eqref{eq:HamDipole}\). Again, we assume \(|z| \approx R\) to approximate the function \(\gamma(|z|) \approx e^2 \Lambda_0^2/(m_e \hbar)\). We now have to write the operator in terms of the holomorphic variables. Since \(z\partial_z = (x+iy)(\partial_x - i\partial_y) = x\partial_x + iy\partial_x - i\partial_y x + y\partial_y\), we can write

\begin{align}
    H_{\mathrm{prec}} = -i\hbar \gamma \sum_j (y_j \partial_{x_j} - x_j \partial_{y_j}) = -\hbar \gamma \sum_j (z_j \partial_{z_j} + z_j^* \partial_{z_j^*})  \,.\label{eq:Hprec}
\end{align}

\noindent If the operator $z_j^* \partial_{z_j^*}$ acts on the wavefunction, it destroys the state since the wavefunction does not depend on $z_j^*$. Therefore, the precession operator $H_{\mathrm{prec}}$~\eqref{eq:Hprec} has mathematically the same form as the confinement potential~\eqref{eq:HconfProject}. We can immediately determine the Hamiltonian to be 

\begin{align}
    \tilde H_{\mathrm{prec}} = -\hbar \Omega^2\frac{\omega_{\mathrm{cyc}}}{\omega_c^2} \frac{\ln(2N_e/\nu)}{N_e} \sum_{l>0} b_l^\dagger b_l\,.
\end{align}

We see that the precession term leads to a reduction in plasmon frequency $\omega_p \to \omega_p - \Omega^2\omega_{\mathrm{cyc}}/(\omega_c^2)\ln(2N_e/\nu)/N_e$. In the limit of ultrastrong plasmon photon coupling, this means $\omega_p \to \omega_p - 2\omega_{\mathrm{cyc}}^2/(\omega_c) \ln(2N_e/\nu)/N_e$. Since neither the cavity frequency nor the cyclotron frequency scale with the system size, the frequency shift will not play a role in the thermodynamic limit, even in the ultrastrong coupling regime. 

Lastly we will consider the second light-matter coupling term $H_{\mathrm{int}} =  -  e\gamma \boldsymbol{\hat e}_z \cdot ( \boldsymbol{r}\times  \boldsymbol{A})$. Using the fact that $(\boldsymbol{r} \times \boldsymbol{\epsilon})_z = ix - y = iz$, we can write for the mode $\boldsymbol{A}_l$ 

\begin{align}
    H_{\mathrm{int, 2}} &=  - \sum_j e\gamma \boldsymbol{\hat e}_z \cdot ( \boldsymbol{r}_j\times  \boldsymbol{A}(\boldsymbol r_j)) = -ie\gamma\sum_{l>0}\sum_j A_l (z_j^l \alpha_l - (z_j^*)^l \alpha_l^\dagger )\\ &= -e\gamma \sum_{l>0} iA_l(p_l \alpha_l - p_l^* \alpha_l^\dagger)\,.
\end{align}

\noindent This interaction term has the same form as Eq. (4) in the main text. This means that the term does not qualitatively change the light-matter coupling but it will change the light-matter coupling strength. Let $\Omega = \Omega_1$ be the light-matter coupling strength considered in the main text. Then the light-matter coupling strength changes to 

\begin{align}
    \Omega_l &\to \Omega/\sqrt{l}\bigg(1 - \frac{\Omega^2 \omega_{\mathrm{cyc}}}{\omega_c^3 }\frac{\ln(2N_e/\nu)}{N_e}\bigg) \ . 
\end{align}

\noindent Again, while the light-matter coupling strength is modified, the modification becomes irrelevant in the thermodynamic limit $N_e \to \infty$. Since we only consider the thermodynamic limit in the main text, we can safely omit these modifications. 

In summary, the terms omitted from the dipole gauge Hamiltonian~\eqref{eq:DipoleGauge} in the main text do not add any new terms to the photon-plasmon Hamiltonian (5). They merely modify the plasmon frequency and the light-matter coupling strength to 

\begin{align}
    \Omega_l &\to \Omega/\sqrt{l}\bigg(1 - \frac{\Omega^2 \omega_{\mathrm{cyc}}}{\omega_c^3 }\frac{\ln(2N_e/\nu)}{N_e}\bigg) \ , \\ 
    \omega_p &\to \omega_p + \frac{\Omega^2}{\omega_c}\frac{\ln(2N_e/\nu)}{N_e}\bigg(1   -\frac{\omega_{\mathrm{cyc}}}{\omega_c}+ \frac{\Omega^2}{\omega_c^3} \omega_{\mathrm{cyc}} \frac{\ln(2N_e/\sqrt{\nu})}{N_e} \bigg) \ .
\end{align}

\noindent However, these contributions vanish in the thermodynamic limit $N_e\to \infty$. This means, for the mesoscopic systems considered in the main text, these terms become irrelevant, i.e., the parameters of our theory are not renormalized.\\

\begin{figure}
    \centering
    \includegraphics[width = 0.6\textwidth]{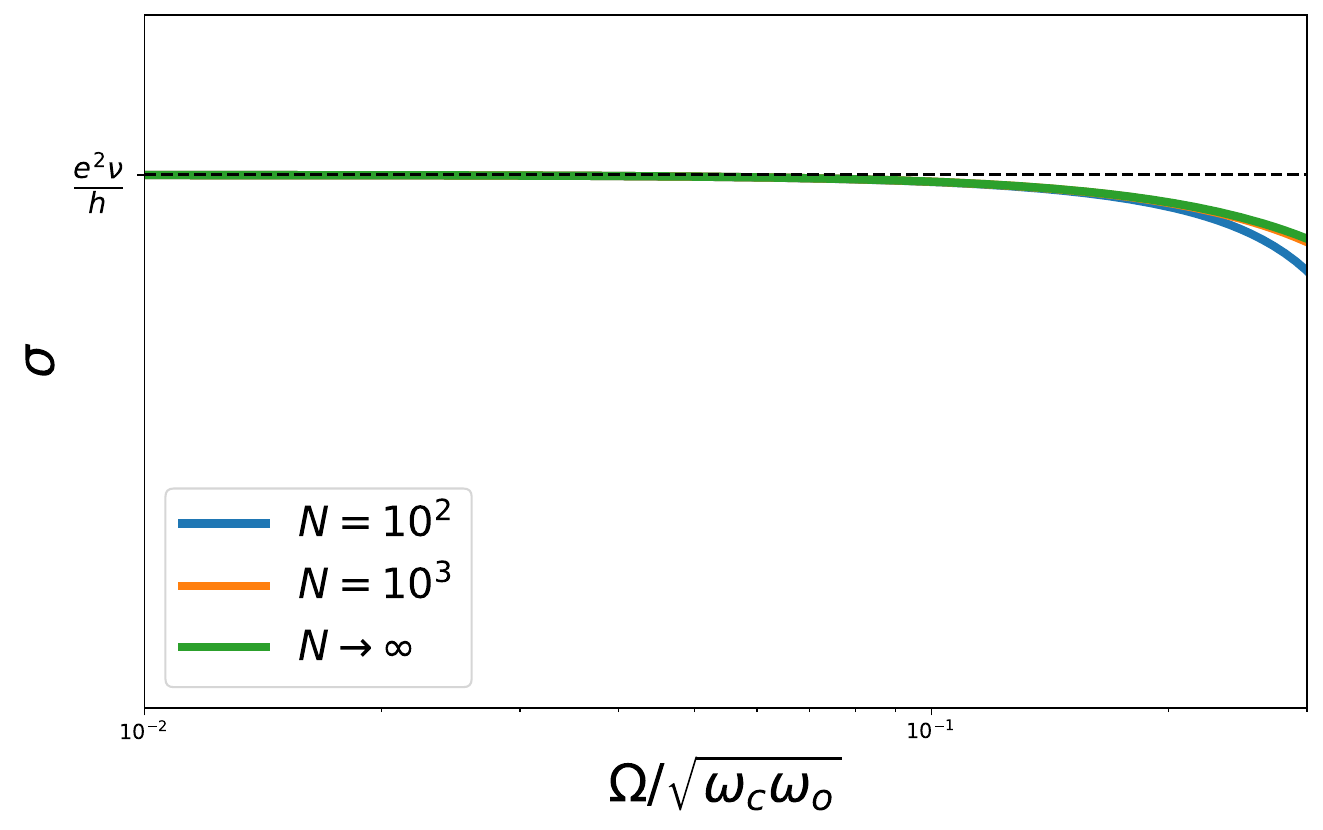}
    \caption{The graph displays Hall conductivity versus normalized light-matter coupling strength for system's with different electron population, \(N\). In the thermodynamic limit (\(N \to \infty\)), the effect of the dipole self-energy disappears, aligning the conductivity results with those presented in the main text. For systems containing \(N = 10^3\) electrons, conductivity closely approximates that of the thermodynamic limit, indicating negligible impact from the dipole self-energy. Conversely, for smaller systems (\(N = 10^2\)), the dipole self-energy significantly disrupts the stability of the Hall plateaus. The parameters used in this plot, chosen for their experimental relevance, include cyclotron frequency \(\omega_{\mathrm{cyc}} = 200\)GHz, coupling frequency \(\omega_c = 0.1 \omega_{\mathrm{cyc}}\), oscillator frequency \(\omega_o = 1\)GHz, and filling factor \(\nu = 3\), as specified in Ref. \cite{ciuti_cavity-mediated_2021}. }
    \label{fig:conductivityFiniteSize}
\end{figure}

To further substantiate the robustness of our findings presented in the main text, we explore the influence of the dipole self-energy term on the topological protection in systems characterized by a finite electron number, denoted by $N$. In Figure~\ref{fig:conductivityFiniteSize}, we depict the conductivity as a function of the normalized light-matter coupling strength, $\Omega/\sqrt{\omega_c \omega_p}$. Notably, as $N$ approaches infinity in the thermodynamic limit, the impact of the dipole self-energy diminishes. Consequently, the green curve aligns with the outcomes illustrated in Figure 3(b) of the main text. For systems with a large electron population ($N > 10^3$), our analysis confirms that the conductivity closely approximates that of the thermodynamic limit, thus validating the assumptions stated in the main text. However, in smaller systems (with $N = 100$), we observe that the dipole self-energy induces a significant deviation of the Hall conductance from its universal value. This finding highlights that the breakdown of topological protection reported in the main text is not merely an artifact of neglecting the dipole self-energy.  

\section{Symmetry constraints to plasmon-photon coupling}
\setcounter{equation}{0}
\label{sec:Symmetry}

In the main text, we argued, based on rotational symmetry, that the total angular momentum of photons and plasmons should be conserved. In this section, we formalize this argument and explicitly show that $H_D$ [cf.~Eq.~\eqref{eq:DipoleGauge}] commutes with the total angular momentum operator $L_z = L_{z,e} + L_{z,p}$, which consists of the electron angular momentum $L_{z,e}$, and the photon angular momentum $L_{z,p}$.  

The photons carry the angular momentum $l = m+1$ composed of one quantum spin angular momentum (SAM) and $m$ quanta of orbital angular momentum (OAM). This means that the cavity electric field $\boldsymbol{E} = i(E(z) \boldsymbol{\epsilon} a - A^*(z)\boldsymbol{\epsilon}^* a^\dagger)$ has a circular polarization vector $\boldsymbol{\epsilon} = (1, i, 0)/\sqrt{2}$, and a mode profile given by $E(z) = z^m$.

For the following considerations, we assume that the system is in the strong coupling regime. In the previous section~\ref{sec:DipoleGauge}, we discussed that the Hamiltonian in the dipole gauge~\eqref{eq:HamDipole} simplifies to Eq.~\eqref{eq:DipoleGaugeStrongCpupling} in the strong coupling regime. Furthermore, we know that both the cavity and the FQH system conserve angular momentum. Consequently, angular momentum conservation can only be broken by the light-matter coupling term in Eq.~\eqref{eq:DipoleGaugeStrongCpupling} 
\begin{align}
    H_{\mathrm{int}} =  \frac e l \ \boldsymbol E_l \cdot \boldsymbol r   \,. \label{eq:Hint}
  \end{align}
\noindent  Here, we substituted the expression for the electric field $\boldsymbol{E}$ and used the fact that the scalar product between the position and the polarization vector $\boldsymbol{\epsilon} = (1, i,0)/\sqrt{2}$, and the electron position $\boldsymbol{r}$ is given by  $\boldsymbol{r}\cdot \boldsymbol{\epsilon} = R z/\sqrt{2}$.

To check whether $H_{\mathrm{int}}$~\eqref{eq:Hint} conserves angular momentum, we need the angular momentum operators for the electrons and the cavity photons. Classically, the angular momentum operator for the electrons in the $z$ direction is given by $L_{z, e} = \boldsymbol{\hat e}_z \cdot (\boldsymbol{r} \times \boldsymbol{p}) = x p_y - y p_x$. Using canonical quantization, $L_{z,e}$ can be extended to the quantum regime by replacing the observables with the corresponding position and momentum operators. This yields the following differential operator $L_{z,e} = -i\hbar ( x \partial_y - y \partial_x)$. Casting the operator in polar coordinates generates the angular momentum operator
\begin{align}
    L_{z,e} = -i\hbar \partial_\varphi\ . \label{eq:Lzelectrons}
\end{align} 
\noindent Thus, we have found the angular momentum operator for the electrons. 

For the photons, we can use the fact that they carry $l = m+1$ quanta of angular momentum. The total angular momentum of the cavity can then be  determined using the number operator
\begin{align}
    L_{z, p} = \hbar la_l^\dagger a_l \ . \label{eq:photonMomentum}
\end{align}

To show angular momentum conservation, we have to show that the total angular momentum operator $L_z = L_{z,e} + L_{z,p}$, consisting of the electron angular momentum $L_{z,e}$~\eqref{eq:Lzelectrons} and photon angular momentum $L_{z,p}$~\eqref{eq:photonMomentum}, commutes with the interaction Hamiltonian $H_{\mathrm{int}}$~\eqref{eq:Hint}. Evaluating the commutator yields 
\begin{align}
    [L_z, H_{\mathrm{int}}] = \bigg[-i\hbar \partial_\varphi, \frac e l\bigg( z^l a - (z^*)^{l} a^\dagger\bigg)\bigg] + \bigg[\hbar l a^\dagger a, \frac e l\bigg( z^l a - (z^*)^{l} a^\dagger\bigg)\bigg]\ . \label{eq:commSymmLz}
\end{align}
\noindent We apply the commutators $[-i\hbar \partial_\varphi, z^l] = \hbar l z^l$. Then, Eq.~\eqref{eq:commSymmLz} becomes 
\begin{align}
    [L_z, H_{\mathrm{int}}] = \hbar  e\bigg( z^l a + (z^*)^{l} a^\dagger\bigg) - \hbar  e\bigg( z^l a + (z^*)^{l} a^\dagger\bigg) = 0\ . \label{eq:commLz}
\end{align}
The commutator~\eqref{eq:commLz} shows that the total angular momentum is conserved by the Hamiltonian $H_D$~\eqref{eq:DipoleGaugeStrongCpupling}. This means that as the edge plasmons carry angular momentum, the light-matter coupling term must have the form $H_{\mathrm{int}} \propto b_l^\dagger a - a^\dagger b_l$ since otherwise the total angular momentum is not conserved. This  determines the shape of the photon plasmon coupling Hamiltonian.

Up to this point, we have assumed that the cavity electric field $\boldsymbol{E}$ has the same chirality as the plasmon modes. For opposing chiralities, one might expect that rotation invariance breaks down and that the angular momentum is no longer conserved. We can, however, obtain the cavity field with opposite chirality by the symmetry transform $x \to x$ and $y \to -y$. This means $\boldsymbol{\epsilon} \to \boldsymbol{\epsilon}^*$ and $E(z) \to E^*(z)$. In this case, the interaction Hamiltonian $H_{\mathrm{int}}$~\eqref{eq:Hint} becomes
\begin{align}
    H_{\mathrm{int}} =  \frac e l \ \boldsymbol E_l \cdot \boldsymbol r  = \frac e l\bigg( (z^*)^{l} a -  z^l a^\dagger\bigg) \ .  \label{eq:Hint2}
\end{align} 
\noindent  By computing the commutator of $H_{\mathrm{int}}$~\eqref{eq:Hint2} with $L_z$ [cf.~Eqs.~\eqref{eq:Lzelectrons} and~\eqref{eq:photonMomentum}], we find 
\begin{align}
    [L_z, H_{\mathrm{int}}] = 2\hbar e\bigg( z^l a + (z^*)^{l} a^\dagger\bigg) \neq 0 \ . \label{eq:commLz2}
\end{align}
\noindent The resulting commutator~\eqref{eq:commLz2} shows that the total angular momentum is no longer conserved by the electron-photon interaction. This also implies that the total particle number of photons and plasmons is not necessarily  conserved any longer. 

While for the opposing chirality of the edge waves and cavity photons the sum of the angular momentum of the FQH system and the cavity is not conserved $L_z = L_{z,e} + L_{z,p}$~\eqref{eq:Lzelectrons}, the difference $\Delta L_z = L_{z,e} - L_{z,p}$ is. Indeed, by evaluating the commutator of $\Delta L_z$ with $H_{\mathrm{int}}$~\eqref{eq:Hint2}, we find 
\begin{align}
    [\Delta L_z, H_{\mathrm{int}}] = \hbar e\bigg( z^l a + (z^*)^{l} a^\dagger\bigg) - \hbar e\bigg( z^l a + (z^*)^{l} a^\dagger\bigg) = 0 \ . \label{eq:commDeltaLz}
\end{align}
\noindent The fact that the angular momentum difference $\Delta L$ between cavity and FQH system is conserved~\eqref{eq:commDeltaLz} also imposes significant constraints on the photon-plasmon interaction Hamiltonian. It means that if a photon-plasmon interaction increases the angular momentum of the cavity by $l$ units, it must also increase the angular momentum on the edge by $l$ units. Therefore, the interaction Hamiltonian consists only of terms of the form $b_l^\dagger a^\dagger$ or $b_l a$. 

In summary, if the chirality of the edge waves matches the cavity mode, the total angular momentum $L_z = L_{z, e} + L_{z,p}$ is conserved~\eqref{eq:commLz}. This implies that the light-matter coupling consists of terms of the form $a^\dagger b_l$ or $a b_l^\dagger$. If the chirality of the light field is opposite to the chirality of the edge wave, the angular momentum difference $\Delta L = L_{z,e} - L_{z,p}$ is conserved~\eqref{eq:commDeltaLz}. Therefore, the photon-plasmon coupling Hamiltonian can only consist of terms of the form $a^\dagger b_l^\dagger$ and $a b_l$.

\section{Light-matter coupling for alternative geometries}

We now compare several sample geometries. Our starting point is the light–matter coupling Hamiltonian
\begin{align}
H_{\mathrm{int}}
= i\hbar\sum_{l\neq 0}
\Big[
 b_l^\dagger\!\left(\Omega_l a_l + \Omega_{-l} a_{-l}^{\dagger}\right)
 + \mathrm{h.c.}
\Big],
\qquad
\Omega_l=\frac{e}{h}\,E_l\,L\,\sqrt{\frac{\nu}{|l|}},
\label{eq:Hmode}
\end{align}
which we will recast in terms of edge field operators. For a single chiral edge of length $L$,
\begin{align}
\phi(x)
&= \frac{\sqrt{\nu L}}{2\pi}\sum_{l>0}\frac{1}{\sqrt{l}}
\Big(e^{-ik_l x} b_l^\dagger + e^{ik_l x} b_l\Big),
\qquad
k_l=\frac{2\pi l}{L},
\label{eq:phi}\\
\rho(x)&=\frac{e}{2\pi}\partial_x\phi(x),
\qquad
j(x)=v\,\rho(x)=\frac{e v}{2\pi}\partial_x\phi(x),
\label{eq:jrho}\\ 
E(x)
&= -\,i\sum_{l>0}\omega_{c,l}A_{0,l}
\Big(e^{ik_l x}a_l - e^{-ik_l x}a_l^\dagger\Big),
\qquad
E_l\equiv \omega_{c,l}A_{0,l}.
\label{eq:E}
\end{align}
Here $E(x)$ is the component of the electric field \emph{along the edge}; polarization has been projected onto the tangent. Using Eq.~\eqref{eq:phi}, the current admits the explicit mode expansion
\begin{align}
j(x)= \frac{e v\sqrt{\nu L}}{(2\pi)^2}
\sum_{l>0}\frac{k_l}{\sqrt{l}}
\Big(e^{ik_l x} b_l - e^{-ik_l x} b_l^\dagger\Big).
\label{eq:jmode}
\end{align}

In the current gauge, $\mathbf{E}=-\partial_t\mathbf{A}$ (cf. the mode choice in Eq.~\eqref{eq:E}), the interaction takes the compact field form
\begin{align}
H_{\mathrm{int}}
= i \int_0^L\!dx\; j(x)\,E(x).
\label{eq:HintField}
\end{align}

We derived Eq.~\eqref{eq:HintField} from first principles for a circular edge in the main text. In what follows we assume the same structure holds for arbitrary edge profiles and use it to analyze different geometries.


We write the interaction in \emph{field} form along a generic edge curve $\Gamma$ with arc–length coordinate $s$ and unit tangent $\hat{\boldsymbol t}(s)$:
\begin{align}
H_{\mathrm{int}}
= i\int_{\Gamma}\!ds\; j(s)\,E_t(s),
\qquad
j(s)=\frac{ev}{2\pi}\,\partial_s\phi(s),
\qquad
E_t(s)=\boldsymbol E(\boldsymbol r(s))\!\cdot\!\hat{\boldsymbol t}(s).
\label{eq:Hint_field_general}
\end{align}
Any \emph{single} boundary component is topologically a circle $S^1$ of length $L$, so its long-wavelength eigenmodes are the Fourier harmonics in $s$, independent of the shape of $\Gamma$:
\begin{align}
\phi(s)
= \frac{\sqrt{\nu L}}{2\pi}\sum_{n>0}\frac{1}{\sqrt{n}}
\Big(e^{-ik_n s} b_n^\dagger + e^{ik_n s} b_n\Big),
\qquad
k_n=\frac{2\pi n}{L}.
\label{eq:phi_general}
\end{align}
Projecting the cavity field onto the edge in the current gauge ($\boldsymbol E=-\partial_t\boldsymbol A$) gives
\begin{align}
E_t(s)
= -\,i\sum_\mu \omega_{c,\mu} A_{0,\mu}
\Big(f_\mu(s)\,a_\mu - f_\mu^*(s)\,a_\mu^\dagger\Big),
\qquad
\mathcal{F}_{n,\mu}\equiv \frac{1}{L}\int_{\Gamma}\!ds\; f_\mu(s)\,e^{ik_n s},
\label{eq:Et_projected}
\end{align}
and defines the $n$th edge-harmonic field amplitude
\begin{align}
E_n \equiv \sum_\mu \omega_{c,\mu} A_{0,\mu}\,\mathcal{F}_{n,\mu}.
\label{eq:En_def}
\end{align}
In terms of photon operators $a_{\pm n}$ associated with these edge harmonics (appropriate linear combinations of the 3D cavity modes $a_\mu$), the light–matter coupling takes the universal form
\begin{align}
H_{\mathrm{int}}
= i\hbar\sum_{n>0}\!
\Big[
 b_n^\dagger\!\left(\Omega_n\,a_n + \Omega_{-n}\,a_{-n}^{\dagger}\right)
 + \mathrm{h.c.}
\Big],
\qquad
\boxed{\ \ \Omega_n=\frac{e}{h}\,L\,\sqrt{\frac{\nu}{n}}\;E_n\ \ }.
\label{eq:Hint_universal}
\end{align}
Thus, plasmon $n$ couples \emph{diagonally} to the $n$th Fourier component of the \emph{tangential} cavity field along the edge, via Eq.~\eqref{eq:Hint_universal}. This selection rule follows from the compact $S^1$ geometry and does not require rotational symmetry.

\paragraph*{Circular disk (single edge).}
Parametrize the circle by $s=R\theta$ ($L=2\pi R$). An OAM-$l$ cavity mode has $f(\theta)\propto e^{il\theta}$ on the 2DEG plane, so
\begin{align}
\mathcal{F}_{n,\mu}=\frac{1}{2\pi}\!\int_{0}^{2\pi}\!d\theta\;e^{i(n-l)\theta}
= \delta_{n,l}.
\label{eq:disk_diagonal}
\end{align}
Substituting Eq.~\eqref{eq:disk_diagonal} into Eq.~\eqref{eq:Hint_universal} reproduces the angular-momentum–diagonal coupling with $n=l$ described in the main text.

\paragraph*{Hall bar (rectangular / elongated).}
A bar has two approximately straight edges at $y=\pm W/2$ and length $L_x$. Let $x\in[0,L_x]$ and $\alpha\in\{\mathrm{top},\mathrm{bot}\}$ label the edges. The interaction reads
\begin{align}
H_{\mathrm{int}}= i\sum_{\alpha}\int_0^{L_x}\!dx\; j_\alpha(x)\,E_{t,\alpha}(x),
\qquad
j_\alpha(x)=\frac{ev}{2\pi}\,\partial_x\phi_\alpha(x),
\label{eq:bar_Hint}
\end{align}
with edge bosons
\begin{align}
\phi_\alpha(x)=\frac{\sqrt{\nu L_x}}{2\pi}
\sum_{n>0}\frac{1}{\sqrt{n}}
\Big(e^{-iq_n x}b_{n,\alpha}^\dagger+e^{iq_n x}b_{n,\alpha}\Big),
\qquad
q_n=\frac{2\pi n}{L_x}.
\label{eq:bar_phi}
\end{align}
Projecting the field on each edge, $E_{t,\alpha}(x)= -i\sum_\mu\omega_{c,\mu}A_{0,\mu}\left(f_{\mu,\alpha}(x)a_\mu-\mathrm{h.c.}\right)$, and defining
\begin{align}
\mathcal{F}_{n,\mu,\alpha}=\frac{1}{L_x}\int_0^{L_x}\!dx\; f_{\mu,\alpha}(x)\,e^{iq_n x},
\qquad
E_{n,\alpha}=\sum_\mu \omega_{c,\mu}A_{0,\mu}\,\mathcal{F}_{n,\mu,\alpha},
\qquad
\Omega_{n,\alpha}=\frac{e}{h}\,L_x\sqrt{\frac{\nu}{n}}\,E_{n,\alpha},
\label{eq:bar_formfactors}
\end{align}
one obtains
\begin{align}
H_{\mathrm{int}}= i\hbar\sum_{n>0}\sum_{\alpha}
\Big[
b_{n,\alpha}^\dagger\!\left(\Omega_{n,\alpha}\,a_n + \Omega_{-n,\alpha}\,a_{-n}^\dagger\right)
+ \mathrm{h.c.}\Big].
\label{eq:bar_Hint_final}
\end{align}
\emph{Implications.} (i) If $E_{t,\alpha}(x)$ is uniform in $x$, only the $n=0$ component is present and, by Kohn’s theorem, DC transport remains unaffected—consistent with Eqs.~\eqref{eq:bar_Hint}–\eqref{eq:bar_Hint_final}. (ii) For a standing wave, $f_{\mu,\alpha}(x)\propto \cos(q_c x)$, the coupling is diagonal in $q$ and resonant when $q_n=\pm q_c$, as seen directly from Eq.~\eqref{eq:bar_formfactors}. (iii) Cross-edge structure matters: if $f_{\mu,\mathrm{top}}(x)=\pm f_{\mu,\mathrm{bot}}(x)$, the symmetric/antisymmetric edge combinations $b_{n,\pm}=(b_{n,\mathrm{top}}\pm b_{n,\mathrm{bot}})/\sqrt2$ are bright/dark in Eq.~\eqref{eq:bar_Hint_final}. Corners or weak $v(x)$ variations only induce subleading mode mixing.

\paragraph*{Corbino (annulus: two concentric edges).}
For outer/inner radii $R_{\mathrm{out/in}}$ and perimeters $L_{\mathrm{out/in}}=2\pi R_{\mathrm{out/in}}$, each boundary supports
\begin{align}
\phi_\alpha(\theta)=\frac{\sqrt{\nu L_\alpha}}{2\pi}
\sum_{l>0}\frac{1}{\sqrt{l}}
\Big(e^{-il\theta}b_{l,\alpha}^\dagger + e^{il\theta}b_{l,\alpha}\Big),
\qquad
\alpha\in\{\mathrm{out},\mathrm{in}\}.
\label{eq:corbino_phi}
\end{align}
Projecting an OAM-$l$ cavity mode onto each edge gives
\begin{align}
\mathcal{F}_{l,\alpha}=\frac{1}{2\pi}\int_0^{2\pi}\!d\theta\; f_\alpha(\theta)\,e^{il\theta},
\qquad
E_{l,\alpha}=\sum_\mu \omega_{c,\mu}A_{0,\mu}\,\mathcal{F}_{l,\alpha},
\qquad
\Omega_{l,\alpha}=\frac{e}{h}\,L_\alpha\sqrt{\frac{\nu}{l}}\,E_{l,\alpha}.
\label{eq:corbino_couplings}
\end{align}
Because the two edges have \emph{opposite orientation} (the tangent reverses), their line-integral contributions in Eq.~\eqref{eq:Hint_field_general} enter with opposite signs. Defining
\begin{align}
b_{l,\pm}=\frac{1}{\mathcal N_\pm}\Big(\Omega_{l,\mathrm{out}}\,b_{l,\mathrm{out}}\ \pm\ \Omega_{l,\mathrm{in}}\,b_{l,\mathrm{in}}\Big),
\label{eq:corbino_bright_dark}
\end{align}
the cavity couples only to the ``bright'' mode $b_{l,+}$ with effective strength $\Omega_l^{\mathrm{bright}}\propto \big|\Omega_{l,\mathrm{out}}-\Omega_{l,\mathrm{in}}\big|$, while the orthogonal combination $b_{l,-}$ is dark (up to weak symmetry breaking). For a wide beam, partial cancellation can occur; for a tight beam, the outer edge typically dominates ($\propto R_\alpha$), as made explicit by Eq.~\eqref{eq:corbino_couplings}.

\paragraph*{What changes across geometries—and what does not.}
\begin{itemize}
\item \textbf{Universal:} the edge is a compact 1D chiral boson. For a single boundary, plasmon frequencies remain $\omega_n = v k_n = v(2\pi n/L)$ (from Eq.~\eqref{eq:phi_general}), independent of the detailed shape of $\Gamma$.
\item \textbf{Selection rule:} plasmon $n$ couples to the $n$th \emph{edge Fourier component} of the tangential cavity field [Eq.~\eqref{eq:Hint_universal}]. For a disk this is “$l=n$ OAM” via Eq.~\eqref{eq:disk_diagonal}; for a bar it is longitudinal momentum $q_n$ [Eqs.~\eqref{eq:bar_phi}–\eqref{eq:bar_formfactors}]; for a Corbino annulus it is $l$ on two edges with opposite signs [Eqs.~\eqref{eq:corbino_couplings}–\eqref{eq:corbino_bright_dark}].
\item \textbf{Mixing:} breaking rotational symmetry does not remove the rule—it only causes a given 3D cavity mode to project onto multiple edge harmonics (several nonzero $\mathcal F_{n,\mu}$ in Eq.~\eqref{eq:Et_projected}).
\item \textbf{Backscattering \& plateaus:} a near-homogeneous component ($n=0$ on a bar or $l=\pm1$ on a disk, as selected by Eqs.~\eqref{eq:bar_formfactors} and \eqref{eq:disk_diagonal}) respects Kohn’s theorem at DC. Multimode or strongly inhomogeneous fields (broad support in $n$ or $l$) mediate inter-edge coupling and can erode quantization, in line with the mechanism analyzed in the main text through Eq.~\eqref{eq:Hint_universal}.
\end{itemize}


   \newpage

    \section{Details on the applied Schrieffer-Wolff transform}
\setcounter{equation}{0}
In this supplement, we discuss how to analyze the cavity-mediated interactions between opposing edges of the $\chi$LL using a Schrieffer-Wolff transform. Initially, the system is described by the Hamiltonian (6) in the main text. In the main text, we use, for brevity, the definition $b_q = \Theta(q) b_{R,q} + \Theta(-q) b_{L,q}$ and $b_q^\dagger = \Theta(q) b_{R,q}^\dagger + \Theta(-q) b_{L,q}^\dagger$, where $b_{q,R}$ represents a plasmon on the right edge and $b_{q,L}$ represents a plasmon on the right edge. Here, we will use the operators $b_{q,L}$ and $b_{q,R}$ for clarity. The Hamiltonian [cf.~Eq.~(6) in the main text] reads in this case  
\begin{align}
H = \sum_{q\neq 0} \bigg[\hbar v|q| \  b_q^\dagger b_q^{\phantom{\dagger}} + \hbar \omega_{c,l} a_{l_q}^\dagger a_{l_q}^{\phantom{\dagger}}
 + i\hbar \Omega (b_q^\dagger \alpha_{q}^{\phantom{\dagger}} - b_q^{\phantom{\dagger}} \alpha_{q}^\dagger+ b_q^\dagger \alpha_{-q}^{\dagger} - b_q^{\phantom{\dagger}} \alpha_{-q}^{\phantom{\dagger}} ) \bigg]   \ . \label{eq:HamSWStart}
\end{align}
To decouple the cavity from the edge modes, we employ the Schrieffer-Wolf transformation to the Hamiltonian. We seek a unitary transformation that when applied, will perturbatively diagonalize the Hamiltonian up to first order. In other words, the goal is to find an operator $S$ such that $[H_0, S] = V$, where $H_0$ is the unperturbed Hamiltonian and $V$ is the perturbation. We can construct the operator $S$ as
\begin{align}
S = \sum_{q\neq 0} i \Omega\left[\frac{1}{\omega_c - vq} \left( \alpha_q b_{q}^\dagger + \alpha_q^\dagger b_{q}\right) +\frac 1 {\omega_c + vq}\left( \alpha_q b_{q}+ \alpha_q^\dagger b_{q}^\dagger \right)\right] \ .
\end{align}
\noindent We can now apply the unitary transformation $H_{\mathrm{eff}} = e^{S} H e^{-S}$. Using the Baker-Campbell-Hausdorff lemma $e^{S} H e^{-S} = H + [S, H] + \frac 1 2 [S,[S, H]] + \dots$ and neglecting any terms of second order or higher, we find 

\begin{align}
H_{\mathrm{eff}}= \sum_{q\neq 0} \left[\hbar \left(v_p - \frac{\Omega^2}{q^2} \frac {2\omega_c} {\omega_c^2 -q^2v^2}\right)|q|b_{q}^\dagger b_q + \hbar  \frac {\Omega^2} {q^2} \frac {\omega_c} {\omega_c^2 - v^2 q^2}|q|(b_q^\dagger b_{-q}^\dagger + b_{q} b_{-q})\right] \ .
\end{align}
\noindent For convenience, as in the main text, we define $V(q) = -\frac {\Omega^2} {|q|} \frac {2\omega_c} {\omega_c^2 - q^2v^2}$, giving us the same form as displayed in Eq.~(7) in the main text.

\section{Details on the Conductivity Calculation}
\setcounter{equation}{0}
Here, we will delve deeper into the process of calculating the conductivity when dealing with a single-mode and a multi-mode cavity within the framework of Luttinger liquid theory for a system described by the effective Hamiltonian [cf.~Eq.~(7) in the main text]. We will specifically focus on a Luttinger liquid situated between two electrodes. In this configuration, the left electrode sets the chemical potential of the right-movers to $V_s$, while the right electrode maintains the chemical potential of the left-movers at zero. Given the electrode position $x_s$, we can derive the following potential~\cite{giuliani_quantum_2005}
\begin{align}
    V(x) = \begin{cases} V_s\cos(\omega t) \qquad &x< x_s \\0 & \mathrm{otherwise}  
 \end{cases} \, .
\end{align}
\noindent Using Kubo's formula, the conductivity in this setup is generally given by~\cite{giuliani_quantum_2005}
\begin{equation}
I = -e \lim_{\omega \to 0} \int_{-\infty}^0 \chi_{j,n}(x-x', \omega) V_s \ dx' \, , \label{eq:current}
\end{equation}
where $\chi$ is the current density response function. Here, we will not take into account internal screening effects, as the two sides of the Hall disc are far apart. The response function itself is given by~\cite{giuliani_quantum_2005}
\begin{equation}
\chi = - \frac {ie} {\hbar L} \int \frac{dq}{2\pi}e^{iq(x-x')} \int_0^\infty \braket{[j_q(t), n_{-q}(0)]} e^{i(\omega + i\eta)t}\,. \qquad\label{eq:response_func}
\end{equation}
\noindent This means that with the substitution $G_j(q,\omega) = \int_0^\infty dt \ \braket{[j_q(t), n_{-q}(0)]} e^{i(\omega + i\eta)t}$, we obtain the same form as Eq.~(8) in the main text:
\begin{align}
    \sigma = \frac{ e^2 }{\hbar L}\int_{-\infty}^{0} dx'\sum_{q\neq 0} e^{iq(x-x')} G_{j}(q, \omega)\, . \label{eq:conductivityApp}
\end{align}

Equation~\eqref{eq:conductivityApp} shows that the conductivity can be calculated as the sum over individual current correlation functions. By inserting the density operator $n_{q} = \sqrt{\frac{L|q|\nu}{2\pi}}(\beta_q + \beta_{-q}^\dagger)$ and the current operator $j_q = \sqrt{\frac{L|q|\nu}{2\pi}}v_p \ \mathrm{sgn}(q)(\beta_q- \beta_{-q}^\dagger)$ into $G_{j, q}$, we find
\begin{align}
G_j(q,\omega) &= i\int_0^\infty dt e^{i(\omega + i\eta)t} \ \braket{[j_q(t), n_{-q}(0)]} \\ &= i\int_0^\infty dt e^{i(\omega + i\eta)t}\frac{Lq}{2\pi} v\nu \braket{[\beta_q(t) + \beta_{-q}^\dagger(t),\ \beta_{-q}(0) - \beta_{q}^\dagger(0)]} \ . 
\end{align}
\noindent As $\beta_q^\dagger(t) = e^{i\omega_q t}\beta_q(0)$, we find
\begin{equation}
G_j(q, \omega) =  \ i\nu\frac{Lvq }{2\pi} e^{2\varphi_q}\bigg( \frac 1 {\omega - v_q |q|} + \frac 1 {\omega + v_{-q}|q|}\bigg). \label{eq:chi_response}
\end{equation}

\noindent We will now consider the case of a single and a multimode cavity separately:

\subsubsection{Single mode cavity}

In the case of a single-mode cavity, only a single plasmon mode couples to the cavity with wavevector $q_c = \frac{2\pi}{L}l$, where $l$ denotes the angular momentum of the cavity mode. This means that the current correlation functions have the following form 
\begin{align}
    \tilde G_j(q_c,\omega) &= 
         i\frac{Lq_c}{2\pi} v \bigg( \frac 1 {\omega - v_{q_c} |q_c|} + \frac 1 {\omega + v_{-q_c}|q_c|}\bigg)\, ,\label{eq:GjsingleMode} \\ G(q,\omega) &=
         i\frac{Lq}{2\pi} v \frac{\omega}{\omega^2 - v^2 q^2} \ . \label{eq:GjsingleModeNormal}
\end{align}
\noindent Here $\tilde G(q_c, \omega)$ \eqref{eq:GjsingleMode} is the cavity modified plasmon correlation function for $q = q_c$, whereas $G_j(q, \omega)$~\eqref{eq:GjsingleModeNormal} is the plasmon correlation function that unaffected by the cavity. We can now calculate the conductivity by inserting $G_j(q,\omega)$~\eqref{eq:GjsingleMode} into the expression for the conductivity [cf.~Eq.~(8) in the main text]. We obtain 
\begin{align}
    \sigma = -\frac{ i e \nu}{\hbar L}\lim_{\omega \to 0} \bigg(\underbrace{\int_{-\infty}^{0} dx'\sum_{q\neq 0} e^{iq(x-x')} G_{j}(q, \omega)\,}_{ = \sigma_0} + \underbrace{\int_{-\infty}^0 \sum_{q = \pm q_c} e^{iq(x-x')} \bigg(\tilde G(q, \omega) - G(q,\omega)\bigg) }_{=\tilde \sigma}\bigg)\  .
\end{align}

\noindent We can now compute the conductivity $\sigma_0$ (similarly to Ref.~\cite{giuliani_quantum_2005}) by moving to the continuum limit 
\begin{align}
    \sigma_0 = \frac{e^2}{\hbar L}\lim_{\omega \to 0} \int_{-\infty}^0 dx' \int dq\ e^{iq(x-x')} \frac{Lq}{2\pi} v \frac{\omega}{\omega^2 - v^2 q^2}    \ .
\end{align}
\noindent We evaluate the integral using the residue theorem $q = \omega/v$. This will result in (see also Ref. \cite{giuliani_quantum_2005})
\begin{align}
    \sigma_0 = \frac{e^2\nu}{h}\ . \label{eq:sigma0}
\end{align}

\noindent Now, we calculate $\tilde \sigma$. In this case, we use $\lim_{\omega \to 0}\tilde G(q, \omega) = \frac{L}{2\pi} v\ \mathrm{sgn}(q)(1/v_{-q} -  1 /v_q) $ and $\lim_{\omega \to 0} G(q,\omega) = 0$ to find 
\begin{align}
\tilde{\sigma} = \frac{e^2\nu}{\hbar L} \frac{L}{2\pi} \int \mathrm{d}x' \left(\frac{1}{v_{-q_c}} - \frac{1}{v_{-q_c}}\right) \cos(q_c(x-x')) =  \frac{e^2\nu} h v \bigg(\frac{1}{v_q} - \frac{1}{v_{-q}}\bigg) \sin(q_c (x-x'))  \, .  \label{eq:sigmaTilde}
\end{align}
We note under general measurement setups $x - x' = L$ implying $\sin(q_c(x-x')) = \sin(2\pi l) =0$. Therefore the conductivity becomes 

\begin{align}
    \sigma = \sigma_0 + \tilde \sigma  = \frac{e^2\nu} h\ . 
\end{align}

\noindent This means the conductivity is unchanged for a single mode cavity. This is valid both if the cavity mode profile is homogeneous $l = 1$ or the cavity mode profile is inhomogeneous $l>1$. In the main text we discuss the homogeneous case to allow for comparison with experiments.

\subsubsection{Multimode cavity}

For this scenario, we can adopt a continuum description for the summation over 
$q$ as seen in Eq.~\eqref{eq:conductivityApp}. It is important to highlight that a minimum wave vector cutoff $\epsilon$, given by $\Omega(q) = gq^2(\epsilon^2 + q^2)$, must be introduced. This ensures that we do not introduce nonphysical poles into the Green's function. Consequently, the conductivity is expressed as

\begin{align}
    \sigma = \frac{ e^2 \nu }{h}\int_{-\infty}^{0} dx'\int dq\  e^{iq(x-x')} \ q v e^{-2\varphi}  \frac{\omega}{\omega^2 - v_q^2 q^2}\, . \label{eq:conductivityAppMM}
\end{align}

\noindent Using the relation obtained from the Bogolubov transform $\mathrm{tanh}(2\varphi_q) = \frac{V}{v + V}$ from the main text we find 

\begin{align}
    e^{2\varphi} = \lim_{q\to 0}\exp\bigg[ \mathrm{arctanh}\bigg(  \frac{V(q)}{v + V(q)}\bigg)\bigg] = \sqrt{1 +  \frac{V(0)} v} = \sqrt{1- \frac{4\Omega^2}{\omega_p \omega_c}}
\end{align}

\noindent In the following, we will linearize the plasmon dispersion around small wavevector $\omega_q = v_qq = vq$ with $\lim_{q\to 0} v_q = v$. We can evaluate the integral over the wavevector $q$ using the residue theorem. As all poles of the integrated are of first order we can write

\begin{align}
\sigma = \frac{e^2\nu}h  \sqrt{1 - \frac{4\Omega^2}{\omega_c \omega_p}}\  \lim_{\omega \to 0}\sum_j  \lim_{q \to q_{s,j}} \bigg(v(q - q_{j})\frac{\omega }{\omega^2 - v_{q}^2 q^2}e^{i q(x-x_s)}  \,\bigg)  .
\end{align}

\noindent We can now tabulate the poles, where the denominator $\omega^2-v_q^2 q^2 = \omega ^2-q^2 v (v-\frac{4 \Omega^2}{\omega_c (q^2+q_0^2)}) = 0$

\begin{table}[H]
    \centering
    \begin{tabular}{c|c|c}
        $q_j$ & $\lim_{\omega \to 0} q_j$ & $\lim_{\omega \to 0}\mathrm{res}(q_j)$ \\ \hline 
        $\pm \sqrt{\frac{A - B}{2v^2 \omega_c}}$ & $\sqrt{\frac{4 \Omega^2}{v \omega c}-q_0^2}$ & 0 \\ \hline 
        $\pm \sqrt{\frac{A + B}{2v^2 \omega_c}}$ & 0 & $\frac 1 {2v}$
    \end{tabular}
    \caption{Poles and residues of $\sigma$~\eqref{eq:conductivityAppMM}}
    \label{tab:my_label}
\end{table}

\noindent This means the integral~\eqref{eq:conductivityAppMM} yields 

\begin{align}
    \sigma = \frac{e^2\nu}{h}  \sqrt{1 - \frac{4 \Omega^2}{\omega_c \omega_p}} \ . 
\end{align}

\noindent For further discussion please refer to the main text.

\subsection{Irrelevance of higher–harmonic corrections to the confinement potential}
\label{sec:irrelevance_higher_harmonics}

Here we show that the results of the main text hold for arbitrary confinement potentials. For this we will explicitly demonstrate that all higher order harmonic corrections to quadratic confinement are irrelevant in the RG sense.  The low–energy dynamics of a single‐branch chiral Luttinger liquid (CLL)—such as the edge of a fractional quantum Hall droplet—are captured by the Euclidean action\footnote{Throughout we set $\hbar = 1$.}
\begin{equation}
S_{0}[\varphi] \;=\; \frac{1}{4\pi\nu}\!\int\!dt\,dx
\Bigl( \partial_{x}\varphi\,\partial_{t}\varphi \;-\;
      v\,(\partial_{x}\varphi)^{2} \Bigr),
\label{eq:S0}
\end{equation}
where $\varphi(x,t)$ is the chiral boson, $\nu$ the filling fraction, and $v$ the edge velocity.  The charge density is
$\rho(x,t) = \partial_x\varphi/2\pi$, which has scaling (engineering) dimension
\begin{equation}
[\rho] = [\partial_x\varphi] = 1 .
\label{eq:rho_dimension}
\end{equation}

\vspace{0.5em}

A smooth electrostatic edge potential $V_{\mathrm{conf}}(x)=\tfrac12 m\omega^{2}x^{2}$ merely renormalises $v$ and the chemical potential.  
It is therefore \emph{marginal} and does not affect the universal long‐distance theory. A generic even potential can be written as
\begin{equation}
\delta V(x) \;=\; \sum_{n\ge 2} g_{2n}\,x^{2n},
\end{equation}
so that the perturbation to the action reads
\begin{equation}
S_{\mathrm{pert}} = \sum_{n\ge 2} g_{2n} \!\int\!dt\,dx\;
x^{2n}\,\rho(x,t).
\label{eq:S_pert}
\end{equation}

\vspace{0.5em}
\noindent
Because $x$ carries dimension $[-x]=-1$ under the chiral rescaling
$(x,t)\mapsto \mathrm{e}^{\ell}(x,t)$ with dynamical exponent $z=1$, the composite operator in~\eqref{eq:S_pert} has
\begin{equation}
\Delta_{2n} \;=\; 2n + [\rho] - 1 \;=\; 2n .
\end{equation}

\vspace{0.5em}

\noindent Standard momentum‐shell renormalisation gives
\begin{equation}
\frac{d g_{2n}}{d\ell} = (1-\Delta_{2n})\,g_{2n}
                       = (1-2n)\,g_{2n},
\end{equation}
so each coupling obeys $g_{2n}(\ell) \propto \mathrm{e}^{(1-2n)\ell}$.
For every $n\ge 2$ we have $1-2n<0$, hence
\begin{equation}
\boxed{\; g_{2n}(\ell)\;\xrightarrow[\ell\to\infty]{}\;0 \qquad (n\ge2)\;} .
\end{equation}
All higher–harmonic terms are therefore \emph{irrelevant} perturbations. The low–energy edge velocity $v$ and thus the single–branch dispersion are fixed solely by the quadratic part $V_{\mathrm{conf}}^{(2)}$; anharmonic corrections only renormalise subleading, non-universal coefficients. Topological degeneracy, fractional statistics and bulk excitation gaps depend on $\nu$ and Coulomb interactions, not on boundary-irrelevant operators. Therefore the global many-body order chosen in the cavity (Laughlin, composite-Fermi sea, etc.) is unchanged by dropping the $(x^{2n},,n\ge 2)$ terms.

\bibliography{Citations}